\documentclass[11pt,a4paper]{article}

\usepackage{a4}          

\usepackage{amsmath}
\usepackage{amssymb}
\usepackage{cite}
\usepackage{array}
\usepackage{multirow}

\usepackage{ifpdf}
\ifpdf
  \usepackage[pdftex, 
    pdftitle={Fluxed M5-Instantons}, 
    pdfauthor={Max Kerstan, Timo Weigand}, 
    pdfsubject={String theory model building},
    bookmarksopen, bookmarksnumbered, bookmarksopenlevel=2]{hyperref}
\fi

\def\hybrid{\topmargin -20pt    \oddsidemargin 0pt
        \headheight 0pt \headsep 0pt
        \textwidth 6.25in       % A4 paper
        \textheight 9 in       % A4 paper
        \marginparwidth .875in
        \parskip 5pt plus 1pt 
          \jot = 1.5ex
   }
\hybrid
\numberwithin{equation}{section}
\numberwithin{table}{section}

\newcommand{\beq}{\begin{equation}}  \newcommand{\eeq}{\end{equation}}
\newcommand{\bal}{\begin{aligned}}   \newcommand{\eal}{\end{aligned}}
\newcommand{\bea}{\begin{eqnarray}}  \newcommand{\eea}{\end{eqnarray}}

\def\ov{\overline}

\newcommand{\nn}{\nonumber}

\newcommand{\cC}{\mathcal{C}}
\newcommand{\cD}{\mathcal{D}}

\newcommand{\cK}{\mathcal{K}}
\newcommand{\cN}{\mathcal{N}}
\newcommand{\cW}{\mathcal{W}}

\newcommand{\cH}{\mathcal{H}}
\newcommand{\cB}{\mathcal{B}}
\newcommand{\cF}{\mathcal{F}}

\newcommand{\cV}{\mathcal{V}}
 
\newcommand{\KK}{\mathcal{K}}

\newcommand{\cZ}{\mathcal{Z}}

\newcommand{\tw}{\text{w}}
\newcommand{\tv}{\text{v}}

\newcommand{\I}{\text{Im}}

\newcommand{\be}{\begin{equation}}
\newcommand{\ee}{\end{equation}}
\newcommand{\mbb}{\mathbb}

\begin{document}

\baselineskip=14pt
\parskip 5pt plus 1pt

\vspace*{-1.5cm}
\begin{flushright}    % Publication numbers
  {\small

  }
\end{flushright}

\vspace{2cm}
\begin{center}        % Main title
  {\LARGE   Fluxed M5-instantons in F-theory  }
\end{center}

\vspace{0.75cm}
\begin{center}        % Authors
 Max Kerstan and  Timo Weigand
\end{center}

\vspace{0.15cm}
\begin{center}        % Institutes
  \emph{ Institut f\"ur Theoretische Physik, Ruprecht-Karls-Universit\"at Heidelberg, \\
             Philosophenweg 19, 69120 Heidelberg, Germany}
             \\[0.15cm]

\end{center}

\vspace{2cm}

%%%%%%%%%%%%%%%%%%%%%%%%%%%%%%%%%%%%%%%%%%%%%%%
%%%%%%%%%%%%%%%%%%%%%%%%%%%%%%%%%%%%%%%%%%%%%%%
%%%%%%%%%%%%%%%%%%%%%%%%%%%%%%%%%%%%%%%%%%%%%%%
%%%%%%%%%%%%%%%%%%%%%%%%%%%%%%%%%%%%%%%%%%%%%%%
%%%%%%%%%%%%%%%%%%%%%%%%%%%%%%%%%%%%%%%%%%%%%%%
%%%%%%%%%%%%%%%%%%%%%%%%%%%%%%%%%%%%%%%%%%%%%%%
%%%%%%%%%%%%%%%%%%%%%%%%%%%%%%%%%%%%%%%%%%%%%%%
%%%%%%%%%%%%%%%%%%%%%%%%%%%%%%%%%%%%%%%%%%%%%%%

\begin{abstract}

We analyse the non-perturbative superpotential due to M5-brane instantons in F-theory compactifications on Calabi-Yau fourfolds.
The M5 partition function  is obtained via holomorphic factorisation by explicitly performing the sum over chiral 3-form fluxes. Comparison with the partition function of fluxed Euclidean D3-brane instantons in Type IIB orientifolds allows us to fix the spin structure on the intermediate Jacobian of the M5-instanton.
We furthermore analyse the contribution of the M5-instanton to the superpotential in the presence of $G_4$ gauge flux, where the superpotential is dressed with matter fields.
We explicitly evaluate the pullback of $G_4$ onto the M5-brane as a measure for the presence of charged instanton zero modes.
This accounts for the M5 charge both under massless $U(1)$s, if present, and under what corresponds in Type II language to geometrically massive $U(1)$s.

\end{abstract}

\clearpage

%%%%%%%%%%%%%%%%%%%%%%%%%%%%%%%%%%%%%%%%%%%%%%%
%%%%%%%%%%%%%%%%%%%%%%%%%%%%%%%%%%%%%%%%%%%%%%%
%%%%%%%%%%%                 %%%%%%%%%%%%%%%%%%%
%%%%%%%%%%%  DOCUMENT BODY  %%%%%%%%%%%%%%%%%%%
%%%%%%%%%%%                 %%%%%%%%%%%%%%%%%%%
%%%%%%%%%%%%%%%%%%%%%%%%%%%%%%%%%%%%%%%%%%%%%%%
%%%%%%%%%%%%%%%%%%%%%%%%%%%%%%%%%%%%%%%%%%%%%%%
%%%%%%%%%%%%%%%%%%%%%%%%%%%%%%%%%%%%%%%%%%%%%%%

\newpage

\tableofcontents

%%%%%%%%%%%%%%%%%%%%%%%%%%%%%%%%%%%%%%%%%%%%%%%%%%%%%%%%%%%%%%%%%%%%%%%%%%%%%%%%%%%%%%%
\section{Introduction}
%%%%%%%%%%%%%%%%%%%%%%%%%%%%%%%%%%%%%%%%%%%%%%%%%%%%%%%%%%%%%%%%%%%%%%%%%%%%%%%%%%%%%

Instanton effects due to Euclidean branes have received widespread interest both in Type II string theory and in F/M-theory  compactifications.
Part of their importance is owed to the fact that they generate perturbatively absent terms in the superpotential of the four-dimensional effective action.
Examples are non-perturbative F-terms involving the K\"ahler moduli in F/M-theory and Type IIB orientifolds \cite{Witten:1996bn}, which are the backbone of recent approaches to moduli stabilisation in the spirit of \cite{Kachru:2003aw}, or non-perturbative couplings in the charged open string sector \cite{Blumenhagen:2006xt,Haack:2006cy,Ibanez:2006da,Florea:2006si}, which play an important role for phenomenology.

The present paper is concerned with instanton effects due to Euclidean M5-branes in F-theory compactifications on a Calabi-Yau fourfold.
One of our aims is to investigate when such an instanton contributes to a superpotential term involving only the geometric string moduli, and when it has to involve matter fields charged under gauge groups appearing on the 7-branes of the theory.
To appreciate some of the problems that occur, let us recall what makes the study of instanton effects in perturbative Type II compactifications tractable and to what extent these properties carry over to F/M-theory. For a review of these techniques and the original literature see e.g. \cite{Blumenhagen:2009qh,Bianchi:2009ij}.

In Type II string theories, D-brane instanton effects can be studied rather explicitly by quantising the excitations of open strings ending on the Euclidean D-brane.
This allows for a detailed understanding in particular of the zero mode spectrum of the instanton, which is crucial in 
determining which superpotential terms the instanton can contribute to or if it contributes at all. 
 
The instanton moduli and their associated zero modes can be treated with techniques very similar to those used in the study of spacetime-filling D-branes. The zero modes arising from open strings with both ends on the instanton are called neutral zero modes, as they are uncharged under the gauge groups of stacks of spacetime-filling  D-branes that may be present. In addition, charged zero modes can appear at the intersection of the instanton with a stack of spacetime-filling D-branes. The charged zero mode structure of an instanton can again be studied by explicitly quantising the open strings stretching between instanton and D-brane stack, taking into account the appearance of gauge flux, if present, in the relevant boundary conditions. 
Instead of quantizing the charged open string sector between the instanton and the spacetime-filling branes, one can look for the appearance of chiral zero modes by considering the transformation properties of the instanton action under the $U(1)$ gauge symmetries on the D-brane stack. These $U(1)$ selection rules imply that an instanton whose classical action carries a certain charge $q$ can contribute at most to terms in the low-energy action involving suitable open string fields $\phi$ of charge $-q$. 

Of particular interest in this paper are Euclidean D3-brane instantons (short E3-instantons) in Type IIB orientifolds. 
As stressed in \cite{Grimm:2011dj}, the computation of the associated superpotential involves a sum over the set of consistent fluxes on the instanton worldvolume. Aspects of E3-instanton flux had previously been discussed in \cite{Collinucci:2008sq}.
Instanton flux has important consequences for the spectrum of charged zero modes because the net chirality of these charged zero modes depends on the gauge flux both on the D-brane stack and on the instanton \cite{Grimm:2011dj}. Among other things, this alleviates the tension between a chiral matter spectrum and moduli stabilisation by D-brane instantons, which was first pointed out in \cite{Blumenhagen:2007sm}. Instanton fluxes can also lift zero modes due to neutral instanton moduli \cite{Bianchi:2011qh}, with great impact on the structure of the superpotential.

In F-theory compactifications, the phenomenologically important instanton effects mentioned above are comparatively under much less control than in the perturbative Type IIB setting. F-theory can be defined as M-theory compactified on an elliptically fibered fourfold in the limit of vanishing fiber volume. The uplift of the E3-instantons of Type IIB discussed above is then given by vertical M5-brane instantons wrapping the elliptic fiber. While the possible superpotential contributions of such M5-instantons have been studied quite extensively starting with the seminal work of Witten \cite{Witten:1996bn}, many questions in particular related to charged zero modes are still quite poorly understood. Let us briefly mention the different types of difficulties that appear in these investigations.

A general problem arises from the fact that the world-volume theory of the M5-brane involves a self-dual 2-form field, whose dynamics do not admit a simple Lagrangian description. While a full covariant action has been known for some time \cite{Pasti:1997gx}, this action includes an additional auxiliary field and does not lend itself easily to a computation of the partition function which controls the low-energy superpotential contribution. Nevertheless, the partition function of the M5-instanton can be elegantly computed by considering first a simpler auxiliary action and then performing holomorphic factorisation \cite{Witten:1996hc,Henningson:1999dm}. This procedure leads to several different candidate partition functions out of which the correct partition function must be picked. Performing this choice directly in M-theory is non-trivial. 

Our first task in this work is to compute the partition function of vertical M5-instantons in a general F-theory compactification, based on the analysis in \cite{Witten:1996hc,Henningson:1999dm}.
The result can be compared, in situations with a Type IIB orientifold limit, with the E3-instanton analogue. 
We establish an explicit correspondence between the sum over gauge fluxes on the E3-instanton as described in \cite{Grimm:2011dj} and the sum over 3-form fluxes in the M5 partition function. This match allows us to identify the correct partition function of the M5-instanton from the set of possible candidates. Our result is in agreement with the modular invariant partition function found for an M5-brane on a flat $T^6$ in \cite{Dolan:1998qk, Gustavsson:2000kr, Gustavsson:2011ur} and with the analysis of \cite{Dijkgraaf:2002ac} for NS5-instantons dual to gravitational instantons in Type IIB. For the recent discussion of such gravitational instantons via M5-instantons see \cite{Grimm:2011sk}.

Viewed microscopically, instanton zero modes arise in F/M-theory from wrapped M2-brane states, which can alternatively be described in terms of (p,q)-string junctions \cite{Blumenhagen:2010ja, Cvetic:2011gp}. While their explicit study in a microscopic picture is difficult due to the lack of an explicit quantised membrane theory, some insights are provided by indirect arguments.  First, the structure of neutral instanton zero modes
can be determined using geometric techniques by studying the deformations of the instanton divisor \cite{Witten:1996bn}, including situations with background fluxes (see e.g. \cite{Kallosh:2005yu,Saulina:2005ve,Kallosh:2005gs,Martucci:2005rb, Bergshoeff:2005yp,Tsimpis:2007sx}).
In such a way, zero mode spectra can be evaluated in concrete examples with rather sophisticated methods \cite{Blumenhagen:2010pv,Cvetic:2010rq,Blumenhagen:2010ed}.
The interplay with world-volume instanton flux has been studied recently in \cite{Bianchi:2011qh} in the language of E3-instantons in strongly coupled Type IIB theory. Other techniques to study instanton zero modes in F-theory include invoking duality with the heterotic string \cite{Donagi:2010pd,Cvetic:2011gp} or anomaly inflow arguments \cite{Cvetic:2011gp}. Earlier work on instanton zero modes in the context of F-theory model builing includes \cite{Marsano:2008py}; the structure of the so-called universal zero modes has been analysed in detail in \cite{Cvetic:2009ah,Cvetic:2010rq}.

The presence of charged zero modes can similarly be detected even without making use of a microscopic theory in terms of M2-branes. To gain some intuition it is beneficial to recall from the theory of E3-instantons in Type IIB orientifolds that the number of charged chiral zero modes is equivalent to a net $U(1)$ charge of the instanton in the presence of gauge fluxes  \cite{Blumenhagen:2006xt,Ibanez:2006da,Florea:2006si}. 
 The same mechanism that is responsible for the gauging of the axions in the E3 worldvolume action is at work also in F/M-theory, as can be verified by a detailed analysis of the gauged supergravity \cite{Grimm:2010ks,Grimm:2011tb}. An important difference to the weakly coupled Type II theories was discussed in \cite{Grimm:2011tb}, though: In generic configurations some $U(1)$s in Type IIB are massive even in absence of gauge flux. These geometrically massive $U(1)$s are not present below the Kaluza-Klein scale in F-theory even though it is possible to formally make them visible by including non-harmonic forms in the dimensional reduction.
 In Type IIB the full set of $U(1)$ selection rules - geometrically massive and massless ones - is necessary to account for all net charged zero modes; this suggests that in F-theory the obvious $U(1)$ selection rules due to geometrically massless $U(1)$s are not sufficient by themselves either. In \cite{Grimm:2011tb} it was in fact suggested, on the basis of the supergravity analysis,  that in F-theory models with a smooth Type IIB limit the selection rules associated with massive  $U(1)$s continue to be present.

In Type IIB the chiral index that counts the net number of charged zero modes can be calculated by integrating the pullback of the 7-brane flux onto the instanton divisor along the mutual intersection curve.
In F/M-theory  gauge flux is described by suitable $G_4$-fluxes. Euclidean M5-branes in the presence of $G_4$-flux are subject to a Freed-Witten anomaly \cite{Freed:1999vc} whenever the pullback of $G_4$ onto the divisor is non-vanishing.
In  \cite{Donagi:2010pd,Marsano:2011nn}  it was argued that non-vanishing $G_4$-flux along the M5 divisor requires the inclusion of extra M2-brane states to cancel the Freed-Witten anomaly. These would be interpreted as charged zero modes, in agreement with the Type II picture of \cite{Blumenhagen:2006xt,Ibanez:2006da,Florea:2006si}. A contribution of the M5-instanton not involving any charged operators is then possible only if the pull-back of $G_4$ onto the instanton world volume vanishes; this serves as a criterion for the presence of charged zero modes and at the same time as a selection rule.
 
An explicit analysis of this Freed-Witten anomaly hinges on recent progress in the explicit description of gauge flux via $G_4$-fluxes in resolved Calabi-Yau fourfolds \cite{Braun:2011zm,Marsano:2011hv, Krause:2011xj,Krause:2012yh,Grimm:2011fx}.
The computation of the pullback of $G_4$ involves a search for 4-cycles on the instanton divisor which host a non-trivial $G_4$-flux. We address this question in the context of globally $U(1)$ restricted Tate models of the form investigated in \cite{Grimm:2010ez}. The general picture emerging from our analysis is as follows:
First, in the presence of massless $U(1)$ gauge group factors, a special type of surface exists for generic complex structure moduli. The integral of the pullback of $G_4$ over this surface within in the instanton computes the net $U(1)$ charge of the instanton in agreement with the supergravity analysis.
In addition to these $U(1)$ selection rules, certain surfaces are present within in the instanton divisor which can be represented as algebraic surfaces only on special loci in complex structure moduli space; we will argue that nonetheless they lead to a non-trivial Freed-Witten constraint. These surfaces are related to the matter surfaces of the F-theory compactification, i.e. the fibrations of $\mathbb P^1$s in the elliptic fiber over curves in the base space where charged matter states are localised. The associated selection rules cannot be determined from massless $U(1)$s in F-theory.

This general analysis can again be compared to perturbative Type IIB results in situations where such a smooth limit exists \cite{Krause:2012yh}. 
The $U(1)$ selection rules associated with the massless $U(1)$s are found to match as expected. The selection rules due to the extra surfaces, on the other hand, encode what appears in Type IIB as the charge under the geometrically massive $U(1)$ symmetries, confirming the expectations of \cite{Grimm:2011tb}.

The remainder of this paper is organised as follows. In section \ref{M5_partition} we discuss the partition function of an M5-instanton employing the idea of holomorphic factorisation as in \cite{Witten:1996hc,Henningson:1999dm}. In section \ref{O1inIIB} the result of this general computation is compared to the instanton partition function of Type IIB E3-brane instantons, which is reviewed in section \ref{fluxedIIBInst}. Amongst other things, the explicit match in section \ref{sec:O(1)Uplift} identifies the correct holomorphic factor of the M5-instanton superpotential.
In section \ref{sec:Inst_plus_flux} we analyse the interplay between M5-instantons and gauge $G_4$-flux in F-theory. After a general outline of the problem in \ref{sec:M2_zero_modes}, section \ref{sec:selection_rules_Ftheory} analyses the Freed-Witten constraint of vanishing $G_4$-flux pullback and discusses the possible types of surfaces that are relevant in detecting this anomaly. In section \ref{sec:selection_rules_IIB} we compare these results to the known selection rules in Type IIB orientifolds. 
Our conclusions and a summary of open questions are contained in section \ref{sec_concl}. Details of the M-theory effective action in the democratic formulation are relegated to the appendix.

%%%%%%%%%%%%%%%%%%%%%%%%%%%%%%%%%%%%%%%%%%%%%%%%%%%%%%%%%%%%%%%%%%%%%%%%%%%%%%%%%%%%%%%%%%%%%%
\section{The partition function of vertical M5-brane instantons}
\label{M5_partition}
%%%%%%%%%%%%%%%%%%%%%%%%%%%%%%%%%%%%%%%%%%%%%%%%%%%%%%%%%%%%%%%%%%%%%%%%%%%%%%%%%%%%

In this section we discuss the partition function of M5-brane instantons in M-theory compactified on a Calabi-Yau fourfold $Y_4$. Although the results of this section are valid for a general instanton, we will be interested mostly in situations where the fourfold $Y_4$ is elliptically fibered, $\pi: Y_4 \rightarrow B_3$, and where the M5-instanton wraps a  
vertical divisor 
\bea
D_M = \pi^{-1} D^{\rm b}_M
\eea
with $D^{\rm b}_M$ a divisor of the base space $B_3$.
In the F-theory limit the three-dimensional effective action of M-theory on an elliptically fibered $Y_4$ is uplifted to an effective action in four dimensions. In this  limit an M5-instanton on a vertical divisor corresponds to a Euclidean D3-brane (dubbed E3-) instanton on the base divisor $D^{\rm b}_M$ \cite{Witten:1996bn}, more precisely to an E3-instanton wrapping the corresponding divisor on the double cover Calabi-Yau threefold $X_3$.
We will furthermore focus on the generation of a non-perturbative superpotential due to the Euclidean M5-brane in the effective action obtained in the F-theory limit. Since we are only analysing the structure of the partition function, we assume that the uncharged zero mode structure indeed allows for the generation of a superpotential. 
Finally, in this section we concentrate entirely on isolated M5-instantons, i.e. we ignore possible intersections of the M5-brane and the discriminant locus of the elliptic fibration.\footnote{More precisely we assume there are no intersections other than with the $I_1$-locus of the discriminant as is necessary for the instanton to generate a superpotential, i.e. to form an $O(1)$ instanton in Type IIB language.} In the presence of such intersections non-trivial constraints due to the appearance of charged zero modes between the instanton and the 7-branes arise, which are the subject of section \ref{sec:Inst_plus_flux}.

The worldvolume of an M5-instanton wrapped on a divisor $D_M$ carries a chiral 2-form potential $\cal B$ with field strength ${\cal H}$. 
In Euclidean signature, the Hodge star operator squares to $-1$, $\ast^2 = -1$, leading to eigenvalues $\pm i$. Chirality of ${\cal B}$ refers to the fact that the field strength ${\cal H}$  is imaginary self-dual, $\ast {\cal H} = i {\cal H}$. Any 3-form $Q$ on the M5-instanton can be split into a chiral and anti-chiral part according to
\be
Q^{\pm} = \frac12 (1 \mp i \ast) Q \,,  \qquad Q = Q^+ + Q^-.
\ee

Due to the chiral nature of ${\cal B}$ the definition of a covariant action involves auxiliary fields, which are necessary to impose self-duality of ${\cal H}$ \cite{Pasti:1997gx}.  One can bypass the associated technical complications by considering instead an action of the form
\bea
\label{M5action1}
S_{\cB} = - 2 \pi \,  \int_{D_M} \big[ ({\cal H} + a \iota^* C_3) \wedge \ast  ({\cal H} + a \iota^* C_3)   + b \,  {\cal H} \wedge \iota^* C_3 \big],
\eea
where ${\cal H}$ is \emph{not} assumed to be self-dual; rather this condition has to be implemented in a suitable manner in all physical expressions as we will review below \cite{Witten:1996hc}. We have chosen the Hodge star operator such that $\int_{D_M} \cH \wedge \ast \cH$ is \emph{negative} definite, which determines the overall sign of the action. 
 
 Before we proceed let us briefly note that the Freed-Witten anomaly \cite{Freed:1999vc} actually rules out cohomologically non-trivial 4-form flux $\iota^* G_4 \neq 0$ on \emph{free} M5-instantons (with $G_4 = d C_3$). This can also be seen by simply considering the equations of motion of $\cB$ from \eqref{M5action1}. The anomaly due to non-trivial flux must be cancelled by the inclusion of suitable M2-branes ending on the instanton cancelling the tadpole for $\cB$. 
 As stressed above, in this section we only consider isolated M5-instantons with  $\iota^* G_4 =0$. We will discuss the inclusion of background flux on the instanton in section \ref{sec:Inst_plus_flux}.

The idea of \cite{Witten:1996hc} to obtain the chiral partition function from the non-chiral action \eqref{M5action1} is to compute the partition function for the full action (\ref{M5action1}) in a fixed $C_3$-form background. This partition function is given by the path integral
\be
\label{Z_B}
\cZ_{\cB} = \int \cD \cB \ e^{- S_{\cB}}.
\ee
To avoid redundancies in the integration one has to omit $\cB$-field configurations which are pure gauge including those which correspond to large gauge transformations. In the absence of $G_4$ fluxes, the large gauge transformations correspond to a shift of ${\cal B}$ by an element of $H^2(D_M, \mathbb R)$.  The case  $ \iota^*G_4 \neq 0$, where this is broken to a discrete subgroup, will not be discussed until section \ref{sec:M2_zero_modes}.
In the situation of trivial fluxes as considered here the path integral thus in essence splits into an integration over the \textit{non-closed} 2-forms and a sum over non-trivial fluxes $\cH_0 \in H^3(D_M, \mathbb{Z})$.
The quantity \eqref{Z_B} takes the form of a sum of terms which are each a product of a factor containing only the self-dual part of the 3-form flux ${\cal H}^+$ and its conjugate defined in terms of ${\cal H}^-$. In \cite{Witten:1996hc}, the partition function for the theory of a chiral 2-form ${\cal B}$ is identified with the chiral part of one particular of these summands. As we will see, it is possible to directly determine this summand by comparison with the Type IIB result for the superpotential of a D3-brane instanton.

One comment is in order: The above analysis only accounts for the bosonic part of the M5-action involving the $\cal B$-field.\footnote{This corresponds, in the language of D3-brane instantons in Type IIB, to the degrees of freedom descending from the instanton gauge field, i.e. the bosonic piece of the Wilson line multiplet.} To fully account for all degrees of freedom one must specify in addition the superembedding of the M5-brane worldvolume and also the fermionic completion involving the superpartner of the ${\cal B}$-field. These parts of the M5-action have been studied e.g. in \cite{Kallosh:2005yu,Saulina:2005ve,Kallosh:2005gs,Martucci:2005rb, Bergshoeff:2005yp,Tsimpis:2007sx}. In addition to the expression (\ref{Z_B}) the full path integral then involves the integral over the bosonic modes from the embedding and the fermionic zero modes. Both are well-known to be crucial to determine if a superpotential is generated. In the sequel we concentrate only on the piece (\ref{Z_B}) of the partition function, assuming implicitly that the pre-requisites for a superpotential contribution, as first specified in \cite{Witten:1996bn}, are met.

In order to identify which of the two factors contains the contribution of $\cH^+$ or $\cH^-$ it is helpful to choose the coefficients in \eqref{M5action1} in such a way that only the anti-self-dual part $C^-$ of $C_3$ couples to $\cH$. This ensures that $\cH^-$ is decoupled from $C_3$, so after performing the factorisation described above the desired chiral partition function encoding the contributing of $\cH^+$ will be the factor involving $C_3$. Expanding $C_3 = C^- + C^+$ in \eqref{M5action1} it is easy to see that $\cH$ couples only to $C^-$ provided $b= -2ai$. The numerical value of the constants $a$ and $b$ can be determined from the condition that the partition function matches the expression obtained for a D3-instanton in the Type IIB limit of F-theory. As we will see  in section \ref{sec:O(1)Uplift} this leads to 
\bea \label{bparameter}
b=-i,   \qquad {\rm i.e.} \qquad  a=1/2.
\eea
Imposing these conditions and suppressing the explicit pullbacks the part of the action that encodes the dynamics of $\cB$ (modulo self-duality) and the coupling to $C_3$ is therefore given by
\be
\label{M5action2}
S_{\cB} = - 2 \pi \int_{D_M}   \big[  \cH \wedge \ast  \cH  -  2i \, \cH \wedge    C^-  +  \frac{1}{4}  C_3 \wedge \ast C_3 \big] .
\ee
In addition one must include also the couplings to the remaining fields of the M-theory background, in particular to the 6-form potential $C_6$ dual to $C_3$. The dimensional reduction of democratic formulation of M-theory involving both $C_3$ and $C_6$ is worked out in appendix \ref{democratic_reduction}. The complete action we consider is
\be
\label{S_M5}
S_{M5} = 2\pi (\mathrm{Vol}_{D_M} + i \int C_6) + S_{\cB} \ .
\ee
However let us stress again at this point that this is still only an auxiliary action as it does not take into account the self-duality of $\cB$. The true physical content of the theory is obtained from the partition function (\ref{Z_B}) after holomorphic factorisation.

One last remark is in order before we proceed with the computation of the partition function.
The gauge invariant field strength appearing in (\ref{M5action1}) is not ${d {\cal B}}$, but the combination 
\bea
 \Lambda = d {\cal B} + \frac12 \iota^* C_3.
\eea 
In general, gauge invariance under a gauge transformation of $C_3$ requires an associated shift of $\cB$ \cite{Pasti:1997gx},
\be \label{deltaC3}
\delta  C_3 = - d \chi, \qquad \delta {\cal B} = \frac12 \iota^* \chi.
\ee
One important special case occurs when the pullback $\iota^* \chi$ to the M5-instanton worldvolume ${\cal W}_{M5}$ is non-vanishing but closed. This happens in particular when the transformation of $C_3$ is due to a gauge transformation of a three-dimensional $U(1)$ gauge symmetry arising from an expansion of $C_3 = A^A \wedge \omega_A$ into harmonic 2-forms $\omega_A$ on $Y_4$. In this case $\delta \iota^* C_3 = 0$ because the gauge shift is purely along the extended three dimensions, where the instanton is pointlike. The corresponding shift of $\cB$ in \eqref{deltaC3} can in this case be undone by a large gauge transformation of $\cB$. However, we will stick to the generic case in the following and return to this special case in section \ref{sec:Inst_plus_flux}.

While the combination $\Lambda$ is thus invariant, the action (\ref{M5action1}) itself is not invariant under a general gauge transformation of $C_3$. Instead under the gauge transformation above one obtains via integration by parts 
\be
\label{trf_S_B}
\delta S_{\cB} = -i\pi \int_{D_M} \chi \wedge G_4, \qquad \quad G_4 = d C_3. 
\ee
This leads to the crucial conclusion \cite{Witten:1996hc} that the associated partition function $\cZ_{\cB}$ must transform as a section of a non-trivial line bundle over the space of $C_3$-form configurations. Nonetheless, the full action $S_{M5}$ in (\ref{S_M5}) of the M5-instanton is invariant under a gauge transformation as considered above. 
This is due to the fact that the dual field $C_6$ transforms as
\be
\label{trf_C_6}
\delta C_6 = \frac12 \chi \wedge G_4
\ee
such that the field strength $G_7 = d C_6 + \frac12 C_3 \wedge G_4$ remains invariant, see also appendix \ref{democratic_reduction} for details.  
The transformation of $C_6$ precisely cancels the variation \eqref{trf_S_B} such that the total action $S_{M5}$ is invariant. Although we have only demonstrated it for the pseudo-action \eqref{S_M5}, this gauge invariance under gauge transformations $\delta C_3 = - d\chi$ is shared by the full covariant action \cite{Pasti:1997gx}.  The relevance of this gauge shift  for M5-instantons in the presence of fluxes was also stressed in the recent \cite{Marsano:2011nn,Grimm:2011sk}.

We now turn to the explicit computation of the gauge invariant partition function
\be
W_{M5} = \int \cD \cB \ e^{-S_{M5}}.
\ee 
The partition function splits into a sum over the saddle points of the action corresponding to a sum over harmonic flux configurations $\cH_0$ and a path integral over fluctuations of the $\cB$-field around these saddle points\footnote{Note that by a gauge transformation of $C_3$ we can assume without loss of generality that $\iota^* C_3$ consists of a sum of harmonic and co-exact pieces. This implies $d\ast \iota^* C_3 = 0$, so the equations of motion obtained from $S_\cB$ indeed lead to harmonic $\cH_0$ for vanishing $G_4$-flux.}. 
This corresponds to the decomposition
\bea
{\cal H} = {\cal H}_0 + d\delta {\cal B}
\eea
in the action (\ref{M5action2}). The partition function becomes
\be
W_{M5} = \sum_{\cH_0} e^{-S_{M5}[\cH_0]} \int \cD \delta \cB \ e^{-S'_{M5}[d \delta \cB, {\cal H}_0]}.
\ee
The prime indicates that in the second factor  ${\cal H}_0$ only appears in the kinetic cross-term $ \int_{D_M} {\cal H}_0 \wedge \ast d (\delta {\cal B})$.
In the semi-classical approximation the action in the exponent is expanded into fluctuations around the given background configuration up to second order. In particular the Hodge star operator splits into a piece $\ast_0$ determined by the background metric and a piece $\hat \ast$ that contains the metric fluctuations. 
This induces a dependence on the fluctuations associated with the embedding of the M5-brane and the bulk geometry.

At zeroth order in fluctuations of the induced metric on the instanton the path integral over fluctuations of the gauge potential $\delta {\cal B}$ decouples from the classical sum over fluxes, as the fluxes are harmonic and self-dual with respect to the background metric,  
\bea
\label{eq:harmonic_background_metric}
d \ast_0 {\cal H}_0 =0.
\eea
Therefore the kinetic  cross-term in $S'_{M5}[d \delta \cB, {\cal H}_0]$ vanishes and the Gaussian 
path integral $\int \cD \delta \cB$  reduces to an overall prefactor of the classical partition function
\be \label{W_cl}
W_{M5}^{cl.} =   \sum_{\cH_0} e^{-S_{M5}[\cH_0]}.
\ee
At the first non-trivial order in the metric fluctuations this decoupling between ${\cal H}_0$ and $\delta{\cal B}$ breaks down because \eqref{eq:harmonic_background_metric} no longer holds when $\ast_0$ is replaced by the fluctuation-dependent Hodge star $\hat \ast$. 
Each classical term in the sum over ${\cal H}_0$ is then weighted by a flux dependent term $\int D\delta \cB \ e^{-S'_{M5}[d \delta \cB, {\cal H}_0]}$. 
These quantum corrections encode the dependence on the K\"ahler and complex structure moduli of the ambient Calabi-Yau as well as the deformation moduli of the instanton divisor. The path integral takes the form of a shifted Gaussian due to the coupling $d\delta {\cal B} \wedge \ast (\cH_0 + i \ast C^- ) $ appearing in $S'_{M5}$, so it is expected to lead to corrections that scale quadratically in the instanton fluxes. The quantum corrections can naturally be seen as the F-theory analogue of the one-loop corrections to the instanton gauge kinetic function in the Type IIB setting, which we will briefly discuss in section \ref{fluxedIIBInst}. 
As in Type II, extra zeroes of the superpotential can in principle arise from this Pfaffian over certain loci in moduli space. In Type IIB these are interpreted as due to vectorlike states of fermionic zero modes whenever the instanton hits another brane. 
Additional corrections to the partition function arise in the presence of non-vanishing $G_4$-flux, which will be the topic of section \ref{sec:M2_zero_modes}.
For a recent discussion of $G_4$-flux induced backreaction effects in the F/M-theory effective action of spacetime-filling 7-branes see \cite{Grimm:2012rg}.

In the remainder of this section we focus on the computation of the classical piece (\ref{W_cl}), i.e. we will ignore the fluctuation-dependent weighting of  $e^{-S_{M5}[\cH_0]}$. For simplicity of notation we drop the subscript in ${\cal H}_0$ and $\ast_0$. 
In fact, reference \cite{Henningson:1999dm} computes the (classical) partition function\footnote{We expect that in some cases there is an F-term condition which actually restricts the partition sum to run over only a sub-lattice of $H^3(D_M, \mbb Z)$, analogous to the well-known F-term condition for the flux on E3-instantons in Type IIB. This would not alter the formal presentation of the results below apart from replacing $H^3(D_M, \mbb Z)$ by the appropriate subspace. Therefore we ignore this subtlety for now and return to it at the end of section \ref{sec:O(1)Uplift} when we have established the correspondence with Type IIB.}
\bea \label{WQ1}
W(Q) = \sum_{\cH \in \ H^3(D_M, \, \mathbb{Z})} e^{ 2 \pi \int \cal H \wedge \ast \cal H } \,   {\rm exp}( {4 \pi i \int {\cal H} \wedge Q} )
\eea
for some $Q \in H^3(D_M,\mathbb R)$.
To understand the result and adapt it to the case of interest here, i.e. allowing for $Q \in H^3(D_M,\mathbb C)$, we recall 
that for a 3-complex dimensional K\"ahler manifold $D_M$ it is always possible to find a symplectic basis $(E_M, F^N)$  of $H^3(D_M, \mathbb{Z})$,
\bea
\label{integer_basis}
&& \int_{D_M} E_M \wedge F^N = \delta_M^N, \\
&&  \int_{D_M} E_M \wedge E_N = 0 =  \int_{D_M} F^M \wedge F^N,  \qquad  M, N = 1, \ldots , \frac12 b^3(D_M), \nonumber
\eea
such that any form $Q \in H^3(D_M, \mathbb{Z})$ with integer periods can be expanded into $(E_M, F^N)$ with integer expansion coefficients.
The Hodge star operator $\ast$ allows us to express, say, $F^M$ in terms of $E_N$ and $\ast E_N$,
\bea
F^N = X^{MN} E_N + Y^{MN}(\ast E_N).
\eea
Using \eqref{integer_basis} it is easy to see that the matrices $X$ and $Y$ are symmetric, and that for our choice of Hodge star operator $Y$ is negative definite.
In terms of the complex matrix
\bea
Z^{MN} = X^{MN} + i Y^{MN}
\eea
one can define a self-dual/ anti-self-dual complex basis of $H^3(D_M)$ as
\bea
E^+_M = - \frac{i}{2} {\I}\,Z_{MN}(F^n - \ov Z^{NP} \, E_P), \quad E^-_M = \frac{i}{2} {\I}\,Z_{MN}(F^N - Z^{NP} \, E_P), \quad
\ast E^\pm_M = \pm i E^\pm_M, \nonumber
\eea
where $  {\I}\,Z_{MN}  = (  {\I}\,Z^{MN})^{-1}$ .
An element $Q \in H^3(D_M, \mathbb C)$ can then be expanded into (anti-) self-dual components as
\bea
Q = Q_+ + Q_- = Q_+^M E^+_M + Q^M_- E^-_M.
\eea
For a real form $Q$ the two terms are related by complex conjugation such that $(Q_-^M)^* = Q_+^M$.

Slightly adapting the result of  \cite{Henningson:1999dm} for the path integral (\ref{WQ1}), one finds\footnote{This holds up to an anomalous prefactor that will cancel in the full quantum computation \cite{Henningson:1999dm} . Note furthermore that in \cite{Henningson:1999dm}  $Q$ is a real form; thus $W(Q)$ is real and the factors appearing in \eqref{WQ} are related by complex conjugation
\be
W(Q) \simeq \sum_{\alpha, \beta}  \Theta [   {   \tiny  \begin{matrix} \alpha   \\  \beta    \end{matrix} } ] (- Z, - Q_+, - Q_-)  \, \,  \ov{  \Theta [   {   \tiny  \begin{matrix} \alpha   \\  \beta    \end{matrix} } ] (- Z, - Q_+, - Q_-) } \ , \quad Q \in H^3(D_M, \ \mathbb{R}).
\ee
However we will need to be more general.}
\bea
\label{WQ}
W(Q) \simeq \sum_{\alpha, \beta}  \Theta [   {   \tiny  \begin{matrix} \alpha   \\  \beta    \end{matrix} } ] (- Z, - Q_+, - Q_-)  \, \,    \Theta [   {   \tiny  \begin{matrix} \alpha   \\  \beta    \end{matrix} } ] ( \ov{Z}, Q_-, Q_+) .
\eea
The theta-functions 
\bea
  \Theta [   {   \tiny  \begin{matrix} \alpha   \\  \beta    \end{matrix} } ] (Z, Q_+, Q_-)  & = & \exp \left[ \frac{\pi}{2} Q_+^M  \I{Z}_{MN}  ( Q_+^N - Q_-^N) \right]  \\
  && \times \sum_{k_M \in \, \mathbb Z} \exp \left[   i \pi ( (k+\alpha)_M Z^{MN} (k+ \alpha)_N + 2 (k + \alpha)_M (Q_+^M + \beta^M) ) \right]   \nn
\eea
involve a particular choice of $ \frac12 h^3(D_M)$-dimensional vectors $\alpha_M, \beta^N$ with entries $0, \frac{1}{2}$ that label a choice of a line bundle on the intermediate Jacobian $H^3(D_M,\mathbb C)/ H^3(D_M, \mathbb Z)$ of the M5-brane. Such line bundles admit a holomorphic section unique up to rescaling, which is the corresponding theta-function \cite{Witten:1996hc}.  If the M5-brane world-volume is a spin manifold, the theta-functions are in one-to-one correspondence with the inequivalent spin-structures on $D_M$.

The general prescription to construct the correct line bundle, and hence to extract the correct summand in \eqref{WQ}, was given in \cite{Witten:1996hc}. However, this procedure is quite difficult to perform explicitly. For M5-branes which correspond to lifts of E3-instantons from Type IIB theory, we will see that one can identify the correct theta-function directly by comparison with the partition function on the IIB side, without going through with the construction of the line bundle.

Finally let us note that the theta-functions appearing in \eqref{WQ} are respectively holomorphic and antiholomorphic when viewed as sections of a line bundle on the intermediate Jacobian endowed with the covariant derivatives
\be
\frac{D}{D Q^M_+} = \frac{\delta}{\delta Q^M_+} - \frac{\pi}{2} Q_{-}^N \I Z_{NM} \ , \qquad \frac{D}{D Q^M_-} = \frac{\delta}{\delta Q^M_-} - \frac{\pi}{2} Q_{+}^N \I Z_{NM} \ ,
\ee
in the sense that
\be
\frac{D}{D Q^M_+} \Theta [   {   \tiny  \begin{matrix} \alpha   \\  \beta    \end{matrix} } ] ( \ov{Z}, Q_-, Q_+) = 0 = \frac{D}{D Q^M_-} \Theta [   {   \tiny  \begin{matrix} \alpha   \\  \beta    \end{matrix} } ] (- Z, - Q_+, - Q_-).
\ee
As $\cH_+$ couples only to $Q_-$ its contribution to the partition function is contained in the anti-holomorphic factor which is annihilated by $\frac{D}{D Q^M_+}$, while the other factor contains the contribution from the anti-self-dual part $\cH_-$.

We are now in a position to apply this result to the action of the M5-instanton
\be
S_{M5} = 2\pi (\mathrm{Vol}_{M5} + i \int C_6) + S_{\cB} \ ,
\ee
with $S_{\cB}$ given by \eqref{M5action2}. The total classical partition function, containing also the undesired contribution from the anti-self-dual part of the field strength $\cH_-$, is then given by
\be
W_{M5}^{tot} = \sum_{\cH \in \ H^3(D_M, \, \mathbb{Z})} e^{-S_{M5}} = e^{-2\pi \left( \mathrm{Vol}_{M5} + i \int C_6 + \frac{b^2}{4}\int C_3\wedge \ast C_3 \right)} W(-i b C_-),
\ee
where $W(Q)$ is defined in \eqref{WQ1}. The classical partition function of the chiral 2-form is then obtained by picking one of the anti-holomorphic factors from the sum in \eqref{WQ} and is given by
\be
\label{W_M5}
W^{cl.}_{M5} =  e^{-2\pi \left( \mathrm{Vol}_{M5} + i \int C_6 \right)} {\cal Z} [   {   \tiny  \begin{matrix} \alpha   \\  \beta    \end{matrix} } ]
\ee
for some choice of $\alpha$ and $\beta$, with
\be
{\cal Z} [   {   \tiny  \begin{matrix} \alpha   \\  \beta    \end{matrix} } ] = \exp \left[ \frac{\pi}{2}b^2 C_-^M \I Z_{MN} (C_-^N - C^N_+) \right] \sum_{k_M \in\  \mathbb{Z}} e^{ i\pi \left( (k+\alpha)_M \ov{Z}^{MN} (k+\alpha)_N + 2(k+\alpha)_M (\beta^M - ib C_-^M) \right)}.
\ee
For future convenience we have reinstated the parameter $b$. In section \ref{sec:O(1)Uplift} we will be able to fix the value both of $b$ and of 
$\alpha_M$ and $\beta^N$ by comparison with the partition function of the Type IIB E3-brane instanton, allowing us to identify the physical partition function of the chiral scalar from the set of candidates.

%%%%%%%%%%%%%%%%%%%%%%%%%%%%%%%%%%%%%%%%%%%%%%%%%%%%%%%%%%%%%%%%%%%%%%%%%%%%%%%%%%%%%%%%%%%%%%%%%%%%%%%%%%
\section{Fluxed $O(1)$ D3-instantons in Type IIB orientifolds} \label{O1inIIB}

In this section we compare the structure of the M5-partition function with the non-perturbative superpotential induced by Euclidean D3-brane instantons in Type IIB orientifolds. We begin  in \ref{fluxedIIBInst} with a recollection of the moduli dependence of the fluxed instanton superpotential as described in \cite{Grimm:2011dj}. This is followed, in section \ref{sec:O(1)Uplift}, by an explicit match of the partition functions.

\subsection{Moduli dependence of the instanton superpotential}
\label{fluxedIIBInst}
%%%%%%%%%%%%%%%%%%%%%%%%%%%%%%%%%%%%%%%%%%%%%%%%%%%%%%%%%%%%%%%%%%%%%%%%%%%%%%%%%%%%%%%%%%%%%%%%%%%%%%%%%%

A Type IIB E3-brane instanton is a Euclidean D3-brane wrapped along a divisor $D_E$ in a Calabi-Yau threefold $X_3$ and pointlike in ${\mathbb R}^{1,3}$. $X_3$ will be the double cover of the base $B_3$ of the elliptically fibered fourfold $Y_4$ defining our F-theory model, $B_3 = X_3 / \sigma$. The corresponding projection is $ p: X_3 \rightarrow B_3$.
In this paper we will focus on $O(1)$ instantons \cite{Argurio:2007qk,Argurio:2007vqa,Bianchi:2007wy,Ibanez:2007rs}, which wrap a single, irreducible divisor $D_E$ that is mapped to itself  as a whole without being pointwise invariant under the orientifold action, i.e. $D_E = D'_E \equiv \sigma^*D_E$.\footnote{If the instanton wraps a non-invariant divisor it is denoted a $U(1)$-instanton, and one must consider the instanton along $D_E$ and the image-instanton along $D_E'$ separately.
The two cases differ qualitatively in the number of neutral zero modes, as in the case of the $O(1)$ instanton two of the universal fermionic zero modes are projected out by the orientifold action (see \cite{Blumenhagen:2009qh} and references therein for details on the zero mode structure).}

As discussed in \cite{Collinucci:2008sq,Grimm:2011dj}, ${ O}(1)$ instantons can carry worldvolume flux with negative parity under the orientifold action ($\tilde{\cF}_E = - \sigma^* \tilde{\cF}_E$). We denote the gauge invariant combination of the world-volume flux and the Kalb-Ramond $B$-field by $\tilde{\cF}_E = 2\pi\alpha' \cF_E - B$. For such a configuration the D-term constraint $\int_{D_E} J \wedge \tilde{\cF}_E = 0$ is trivially satisfied and does not constrain the world-volume flux. However, an extra constraint arises from the Freed-Witten quantisation condition on the gauge flux
 \bea
 {\tilde \cF}_E + \iota^* B + \frac{1}{2}c_1(K_{D_E}) \in H^2(D_E, \mathbb Z).
 \label{FWanom}
 \eea
 For an $O(1)$ instanton the canonical class is even with respect to the orientifold projection. Thus the orientifold even part of the $B$-field must cancel the half-integer contribution for a non-spin divisor as $\tilde {\cF}_E$ is odd under the orientifold action.

The action of an $O(1)$ E3-instanton on the divisor $D_E \subset X_3$ depends on the flux $\tilde{\cF}_E$ as well as the chiral fields \cite{Grimm:2004uq,Jockers:2004yj}
\bea
G^a &=& c^{a} - \tau b^a, \qquad \qquad a = 1,..., h^{1,1}_- (X_3) \label{GaTalpha} \\
T_{\alpha} &=& \frac{1}{2}\cK_{\alpha \beta \gamma} v^\beta v^\gamma+ i \left(c_{\alpha} - \cK_{\alpha bc} c^b {b}^c \right) + \frac{i}{2} \tau \cK_{\alpha b c}  b^b   \, b^c \;, \qquad \qquad \alpha = 1,..., h^{1,1}_+ (X_3), \nonumber
\eea
where $\tau = C_0 + i\, e^{-\phi}$ represents the axio-dilaton. The moduli $c^a$, $b^a$ and $v^\alpha$ arise from the expansion of the fields $C_2$, $B$ and $J$ into the bases $\{ \omega_\alpha \}$ and $\{\omega_a\}$ of the orientifold even and odd cohomology groups $H^{1,1}_+ (X_3, \mbb Z)$ and $H^{1,1}_- (X_3, \mbb Z)$, respectively. $c_\alpha$ in turn is given by $C_4 = c_\alpha \tilde{\omega}^\alpha + ...$, where $\{ \tilde{\omega}^\alpha \}$ is the basis of $H^{2,2}_+ (X_3, \mbb Z)$ normalised by
\be
\int_{X_3} \omega_\alpha \wedge \tilde{\omega}^\beta = \delta^\beta_\alpha.
\ee
Similarly one has a basis $\{ \tilde{\omega}^a \}$ of $H^{2,2}_- (X_3, \mbb Z)$ which is dual to the $\{\omega_a\}$.
Finally we have also used the triple intersection numbers, written as usual in the supergravity limit as
\be
\cK_{ABC} = \int_{X_3} \omega_A \wedge \omega_B \wedge \omega_C  \in 2\mbb Z, \qquad \qquad A,\ B,\ C = a, \alpha.
\ee
Note that these triple intersection numbers are even integers due to the fact that we have chosen bases with definite orientifold parity.

In order to write down the instanton action it is helpful to decompose the instanton flux as 
\be
\label{exp_F_E}
2\pi\alpha'{\cF}_{E} =  \cF^{a}_{E} \iota^* \omega_a + {\cal F}_E^{\tv},
\ee
where ${\cal F}_E^{\tv}$ denotes the variable flux lying in the orthogonal complement of $\iota^* H^2(X_3)$ in $H^2(D_E)$\footnote{Variable flux ${\cal F}_E^{\tv}$ is flux lying in the kernel of the Gysin map defined on $H^2(D_E)$. Viewed more geometrically, the existence of such flux requires the existence of non-trivial 2-cycles on the instanton divisor that are trivial as 2-cycles in the ambient space $X_3$.}. For a $U(1)$ instanton this expansion would include also positive parity fluxes.\\ 
When computing the action of a D-brane or instanton one must take care of a factor of 1/2 difference between the descriptions on the base $B_3$ and the double cover $X_3$. For a general brane and image brane pair with $D_A \neq D_A'$ one defines the combinations $D_A^\pm = D_A \cup (\pm D_A')$, where the minus sign denotes a reversal of orientation, and introduces the wrapping numbers
\be
C^\alpha_A = \int_{D^+_A} \tilde{\omega}^\alpha , \qquad\quad C^a_A = \int_{D_A^-} \tilde{\omega}^a.
\ee
In terms of the physical theory defined on $B_3$ the brane and image brane are descriptions of the same object, so the physical action is obtained by adding the actions of the brane stacks on $D_A$ and $D_A'$ and dividing by 2. In the case of an invariant object $D_E = D_E'$ such as our $O(1)$ instanton one defines $D_E^+ = D_E$, which implies
\be
[D_E] = C^\alpha_E \omega_\alpha \quad \text{ in } X_3
\ee
for the $O(1)$ instanton. Nevertheless the physical action is still obtained by computing the usual brane action on $D_E$ in $X_3$, \emph{and then dividing the result by 2}. This takes into account the fact that on the double cover all intersection numbers are twice those computed on the base.\footnote{While it may seem unnatural to include it for a single invariant brane on $D_E$, this factor of 1/2 can also be seen to be necessary e.g. for consistency upon recombination of a $U(1)$ instanton-image-instanton pair into an $O(1)$ instanton.}

For a given configuration of instanton flux, the physical action of the $O(1)$-instanton is  \cite{Grimm:2011dj}
\bea
\label{gaugekina}
 S_E   &=&  \pi  \Big( C^\alpha_E ( T_{\alpha} + i  \Delta_{E \alpha}) 
                     +    i  \Delta^{\tv}_{E}      \Big),
                     \eea
                     with
                        \bea             
\Delta_{E \alpha} =    \cK_{\alpha b c} \, G^b \, { \cal F}^c_E  + \frac{\tau}{2} \kappa_{\alpha b c}   { \cal F}^b_E \,   { \cal F}^c_E, \qquad   
\Delta_{E}^{\tv} =  \tau \, \int_{D_E} {\cal F}_E^\tv \wedge  {\cal F}_E^\tv.
\eea
Note that the universal contribution of the $b^a$-moduli is already encapsulated in $T^\alpha$ and $G^a$ given in (\ref{GaTalpha}) so that only the actual gauge flux  $\cF_E$ (as opposed to $\tilde \cF_E$ ) appears in $\Delta_{E \alpha}$.

If the instanton divisor $D_E$ intersects any divisor $D_A$ wrapped by one of the spacetime-filling 7-branes, the instanton flux can lead to a chiral spectrum of charged zero modes between the instanton and the 7-brane stack or, equivalently, to a non-trivial instanton charge under the diagonal $U(1)_A$ realised on the brane stack. 
For the time being we ignore this effect in the same manner as we ignored the interplay between the 7-branes and the M5-instanton in the F/M-theory formulation.
We will turn to this question in section \ref{sec:Inst_plus_flux}.

As in M-theory the full partition function involves a sum over admissible fluxes ${\cal F}_E$ - the classical partition function - together with a path integral over the fluctuations $\delta A$ of the instanton gauge field configuration around these saddle points of the action. Here we will compute only the classical part of the partition function which will be matched with the corresponding results for the M5-instanton obtained in the previous section. However, let us first briefly comment on the structure of the quantum contributions to the partition function from the Type IIB perspective.

The discussion around (\ref{W_cl}) applies in an analogous manner after replacing the 2-form $\cal B$ by the instanton gauge potential and the 3-form strength ${\cal H}$ by the instanton flux ${\cal F}_E$. In this supergravity picture not only the dependence on the Calabi-Yau moduli, but also on the brane moduli associated with the D3- and D7-branes present in the compactification is encoded via their backreaction on the metric. 
Specifically, the corrections to the holomorphic piece of the instanton partition function (which is relevant for the superpotential) induced in this manner by D3-branes and by fluxed D7-branes have been discussed in \cite{Baumann:2006th} and \cite{Marchesano:2009rz} (see \cite{Koerber:2008sx,Baumann:2010sx} for more general aspects of this moduli dependence of the instanton superpotential in terms of the supergravity backreaction). 
These articles consider E3-instantons with vanishing instanton flux. From the IIB version of the discussion around (\ref{W_cl}) (i.e. after replacing $\cal H$ by ${\cal F}_E$) it is clear that the corrections will in general depend quadratically on the instanton flux. This is also in qualitative agreement with \cite{Marchesano:2009rz}, which argues for a correction to unfluxed instantons that depends on the induced D3-charge of fluxed D7-branes wrapping a distant brane divisor $S$. This induced D3-charge is of course quadratic in the gauge fluxes on the D7-brane. By symmetry this in turn suggests that corrections in the presence of suitable D7-branes should also have quadratic dependance on the instanton flux.

Finally recall that the quantum corrections to the instanton partition function can also be determined by a suitable stringy calculation of correlation functions in an instanton background. The backreaction of the instanton and other D-branes on the background geometry corresponds directly to corrections induced by  tree-level diagrams for closed strings interacting with the instanton. These admit an alternative description as one-loop open string diagrams of either Moebius strip topology with only a boundary on the instanton or annulus diagrams with boundaries on the instanton and other D-branes. Indeed such open string one-loop amplitudes are known to describe the instanton 1-loop Pfaffian \cite{Blumenhagen:2006xt,Akerblom:2006hx}. The relevant annulus diagrams that contribute to the holomorphic instanton superpotential have been identified in \cite{Akerblom:2007uc,Billo:2007sw,Billo:2007py}. These are equivalent to the corrections to the holomorphic gauge kinetic function \cite{Berg:2004ek}. Again these results suggest a correction that depends quadratically on the instanton flux.

With this in mind we now turn to the classical partition function function
\be
\label{ZE3}
W^{cl.}_{E3} = \sum_{\cF_E} e^{- S_E [\cF_E]},
\ee
which is the direct analogue of the M5-instanton expression (\ref{W_cl}). 
 It is helpful to rewrite this classical partition function
 in a slightly different manner such that in particular the integer quantisation of the fluxes required by the Freed-Witten condition \eqref{FWanom} is manifest.
To this end we must take into account a minor subtlety related to the expansion of the pullback fluxes in \eqref{exp_F_E}. Note that the forms $\omega_a$ were chosen as an integral basis of $H^{1,1}_- (X_3, \mbb Z)$ such that an integral form on $X_3$ could be expanded into this basis with integer coefficients. However, for general $C^\alpha_E$ the corresponding statement does not hold for the pullbacks $\iota^* \omega_a$ as a generating system for $\iota^* H^{(1,1)}_- (X_3) \cap H^{(1,1)}_- (D_E, \mathbb Z)$, i.e. the $\cF^a_E$ in \eqref{exp_F_E} are \emph{not} necessarily integers even though the flux itself is an integer form.

To facilitate comparison with the M5-instanton partition function we would like to choose a basis $\{ \omega_m \}$ of $\iota^* H^{(1,1)}_- (X_3) \cap H^{(1,1)}_- (D_E, \mathbb Z)$ in which the expansion coefficients are integer. At this point we must again take into account a subtle factor of 2 arising from the process of orientifolding. The Freed-Witten anomaly for the physical theory of unoriented open strings on the orientifold quotient requires the integral of the flux over any 2-cycle 
on the physical instanton divisor on the orientifold quotient of $X_3$
 to be integer. This implies that over the uplift of this curve on the double cover the integral of $2\pi\alpha' \cF_E$ is an \emph{even} integer. Hence if we want $2\pi\alpha' \cF_E \equiv \cF^m \omega_m$ with integer expansion coefficients $\cF^m \in \mbb Z$ we must require on the double cover $X_3$
\be
\label{def_omega_m}
\int_{D_E} \omega_m \wedge \omega_n = 2 \delta_{mn}.
\ee

The new basis forms $\omega_m$ can be related to the pullbacks (which we suppress in the following) of the $\omega_a$ by some matrix $M_a^m$
\be
\label{omega_a_omega_m}
\omega_a = M_a^m \omega_m.
\ee
The matrix $M$ characterises the embedding of the divisor $[D_E] = C^\alpha_E \, \omega_\alpha$ wrapped by the $O(1)$-instanton, and satisfies the useful relation
\be
\label{M_and_C}
C^\alpha_E \,  \cK_{\alpha a b} = \int_{D_E} \omega_a \wedge \omega_b = 2 M_a^m M_b^n \delta_{mn}.
\ee
To define a full basis of $H^{1,1}_- (D_E, \mbb Z)$ we extend this by forms $\omega_{\hat m}$ spanning the orthogonal complement of $\iota^* H^{(1,1)}_- (X_3)$. This allows for an expansion of the variable fluxes in \eqref{exp_F_E} as well. The forms $\omega_{\hat m}$ are by definition orthogonal to the $\omega_m$ and are also taken to fulfill
\be
\label{def_omega_hat_m}
\int_{D_E} \omega_{\hat m} \wedge \omega_{\hat n} = 2 \delta_{\hat m \hat n}.
\ee

An integer quantised flux $2\pi\alpha' \cF_E$ can then be expanded into this basis as 
\be
2\pi\alpha' \cF_E = \cF^a M_a^m \omega_m + (\cF^{\tv})^{ \hat m} \omega_{\hat m} \equiv \cF^M \omega_M \ , \qquad \cF^M \in \ \mathbb{Z} \ .
\ee
Here and in the following the index $M$ runs over the full basis $M = m,\ \hat m$.
Similarly, one has
\be
G = G^a \omega_a = G^a M^m_a \omega_m \equiv G^m \omega_m.
\ee
With the help of these relations we may rewrite the classical partition function of the $O(1)$-instanton \eqref{ZE3} as
\bea
\label{ZE3_rewrite}
W^{cl.}_{E3} &= & \exp \left[-\pi \left( \frac12 C^\alpha_E \cK_{\alpha \beta\gamma}v^\beta v^\gamma + iC^\alpha_E(c_\alpha - \frac12 \cK_{\alpha ab}c^a b^b)\right)\right] \nn \\ 
&& \times  \exp \left[ -\frac{i\pi}{\tau - \ov{\tau}}\delta_{mn}G^m(G^n-\ov{G}^n) \right] \sum_{\cF^M \in\ \mathbb{Z}} e^{-i\pi \left(  2 \delta_{mn} G^m \cF^n + \tau \delta_{MN}\cF^M\cF^N  \right) }.
\eea

\subsection{Matching the M5 and E3 partition functions}
\label{sec:O(1)Uplift}

To match the partition functions of the M5- and E3-instantons we need to match the sum over 3-form fluxes ${\cal H} \in H^3(D_M)$ in M-theory to the sum over admissible $O(1)$ instanton fluxes ${\cal F}_E \in H^{1,1}_-(D_E)$ on the IIB side. An important ingredient is the relation between the two cohomology groups in which the fluxes take their value. 
Recall that the M5-instanton wraps a divisor $D_M$ of the F-theory fourfold $Y_4: T^2\hookrightarrow Y_4 \rightarrow B_3$ which is itself an elliptic fibration $D_M: T^2 \hookrightarrow D_M \rightarrow D_M^{\rm b}$. 
The Type IIB orientifold is defined on the  Calabi-Yau threefold $X_3$ obtained as the double cover $p: X_3 \rightarrow B$ of the F-theory base $B$.
The divisor $D_E \subset X_3$ wrapped by the Type IIB instanton is the preimage $D_E = p^*(D_M^{\rm b})$ of $D_M^{\rm b}$.

For simplicity we only consider Type IIB instantons which do not admit non-trivial Wilson lines, such that $h^1(D_E) = h^3(D_E) = 0$. In this case generic elements of $H^3(D_M)$ arise from 2-forms on $D_E$ with negative orientifold parity. Given a 2-form $\omega \in H^2_-(D_E)$ one can schematically see the corresponding 3-forms as arising from the wedge product with one-forms $dx,\ dy \in H^1(T_2)$ on the elliptic fiber \cite{Blumenhagen:2010ja} (see also \cite{Donagi:2010pd}),
\be
\label{uplift_neg_form}
\omega  \rightarrow   \omega \wedge dx, \qquad \omega  \rightarrow   \omega \wedge dy .
\ee
The negative monodromy of $dx$ and $dy$ around the orientifold plane cancels the negative parity of $\omega$ such that the product gives rise to a well-defined 3-form on $D_M$. This qualitative picture was confirmed in an an explicit example in \cite{Blumenhagen:2010ja}, where it was found that
$H^{(3,0)}(D_M)$ is the uplift of  $H^{(2,0)}_-(D_E)$, while 
 $H^{(2,1)}(D_M)$ receives contributions from the uplift of $H^{(2,0)}_-(D_E)$ and $H^{(1,1)}_-(D_E)$. We will denote the two subspaces of $H^{(2,1)}(D_M)$ related  to $H^{(2,0)}_-(D_E)$ and to $H^{(1,1)}_-(D_E)$ by $H^{(2,1)}_{\rm hor}(D_M)$ and $H^{(2,1)}_{\rm ver}(D_M)$, respectively. Finally, $H^{(2,1)}(D_M)$ receives a further contribution from negative parity one-forms on the intersection of the instanton with D7-branes, counted by $h^1_-(D_E \cap D7)$. We will always assume that this possible contribution is absent, $h^1_-(D_E \cap D7)=0$ \cite{Blumenhagen:2010ja}.

There is an important caveat to the above relation between $H^{(2,1)}_{\rm ver}(D_M)$ and $H^{(1,1)}_-(D_E)$ that follows from the analysis in \cite{Grimm:2011tb}:
The argument \eqref{uplift_neg_form} for the uplift of the latter fails if $\omega$ has non-vanishing pullback onto an intersection of the instanton with a D7-brane on a divisor $D_A$ above which the elliptic fiber degenerates. As $D_E \cap D_A$ is non-trivial as a 2-cycle in the ambient space $X_3$ this is possible for forms in $\iota^* H^{(1,1)}_-(X_3) \cap H^{(1,1)}_-(D_E) $ which arise as pullbacks from the ambient space. In the language of section \ref{fluxedIIBInst} the (1,1)-forms for which the uplift \eqref{uplift_neg_form} can fail to lead to an element of $H^3(D_M)$ are therefore the forms $\iota^* \omega_a$ with
\be
\label{pullback_brane_instanton}
0 \neq \int_{D_E\cap D_A} \iota^* \omega_a = \frac12 \cK_{a b \alpha} C^b_A C^\alpha_E, \qquad \text{ where } [D_E] = C^\alpha_E \omega_\alpha, \ [D_A] = \frac12 (C^\alpha_A \omega_\alpha + C^a_A \omega_a).
\ee
Note that this is not in contradiction with the results of \cite{Blumenhagen:2010ja}, as in the example considered there $H^{(1,1)}_-(X_3) = 0$.

By the discussion above we see that the uplift of $H^{(1,1)}_-(D_E)$ can become problematic only if there are D7-branes with odd wrapping numbers $C^a_A \neq 0$ present in the model. It was argued in \cite{Grimm:2011tb} that the diagonal $U(1)_A$ of the D7-brane stack, which is massive by the geometric St\"uckelberg mechanism, in this situation is described in M-theory by certain non-closed forms. 
To circumvent this subtlety we assume for now that $C^a_A = 0$ for all D7-brane stacks present, such that no geometric gauging occurs. In particular we therefore have an uplift
\be
\label{uplift_h_11}
H^{(1,1)}_-(D_E) \rightarrow H^{(2,1)}_{\rm ver}(D_M).
\ee
We will comment on the more general case at the end of section \ref{sec:selection_rules_IIB}.

We are now in a position to explicitly match the superpotential contributions generated by the E3- and M5-instantons. Let us first assume that the E3-instanton is wrapped on a rigid divisor, i.e. $H^{(2,0)}_-(D_E) = 0$. In light of the preceding discussion we then have an isomorphism between $H^{(2,1)}(D_M)$ and  $H^{(1,1)}_-(D_E)$. Note that the divisor $D_M^{(\rm b)}$ in the base $B_3$ will intersect  the discriminant locus $\Delta$ at least along the piece of $I_1$-locus which is the uplift of the orientifold plane, as is necessary for the instanton to be of $O(1)$-type. It can also intersect further components of the discriminant locus corresponding to stacks of D7-branes. However, as we have assumed that all branes are wrapped on orientifold-even divisors the negative parity 2-forms in $H^{(1,1)}_-(D_E)$ have no support on the O7-plane and the D7-branes. Hence we expect the uplift \eqref{uplift_h_11} to hold despite the fact that the elliptic fibration degenerates over $\Delta \cap D_M^{(\rm b)}$.

As in section \ref{fluxedIIBInst} we must take care regarding the quantisation condition of $\cH$-flux along forms that arise by pullback of the ambient space basis $\alpha_a, \beta^a \in H^3(Y_4)$. This basis is the uplift of the basis $\omega_a$ of $H^2_-(B_3)$ according to the prescription (\ref{uplift_neg_form}). 
As before these pullbacks do not necessarily form an \emph{integral} basis of $H^3(D_M, \mathbb{Z}) \cap \iota^* H^3(Y_4)$. Hence they cannot immediately be identified with a subset of the 3-forms $E_M$, $F^N$ introduced in (\ref{integer_basis}) as an integral basis of $H^3(D_M, \mathbb Z)$. One expects that the $E_M$, $F^M$ arise in a similar manner to \eqref{uplift_neg_form} from the forms $\omega_M$ introduced in \eqref{def_omega_m} resp. \eqref{def_omega_hat_m}. Hence we can similarly split $M = m,\ \hat m$, where $E_m$, $F^m$ span the space of forms obtained by pullback while $E_{\hat m}$, $F^{\hat n}$ span the orthogonal complement. In full analogy to \eqref{omega_a_omega_m} one also gets
\be
\iota^* \alpha_a = M_a^m E_m, \qquad\quad \iota^* \delta_{ab}\beta^b = M_a^m \delta_{mn} F^n.
\ee
Note that this is consistent with the symplectic structure of \eqref{integer_basis} and \eqref{M_and_C} if one uses the intersection properties of the $\alpha_a$ given in \eqref{intersection_numbers_fourfold}.

Now comparing the M5-instanton partition function \eqref{W_M5} with the partition function of the E3-instanton given in \eqref{ZE3_rewrite} one finds a perfect match with the help of the identifications\footnote{Note that of course $C_\pm$ has expansion coefficients only along the forms $E_m$, $F^m$ as it arises by pullback, or in other words $C_\pm^{\hat m} = 0$. We should also point out that the diagonal form of $Z^{MN}$ is a consequence of the chosen basis of forms $\omega_M$ giving rise to $E_M$, $F^M$, which were taken to obey \eqref{def_omega_m} and \eqref{def_omega_hat_m}.} 
\be
\label{identifications_for_partition_function}
\cF^M \leftrightarrow \delta^{MN} k_N \ ; \quad G^m \leftrightarrow  C_-^m \ ; \quad \ov{G}^m \leftrightarrow  C_+^m \ ; \quad \ov{Z}^{MN} = - \tau \delta^{MN}.
\ee
 In order to obtain this match two non-trivial assertions must be made.
Firstly, comparison with Type IIB fixes the parameter 
$b = -i$ in the M5-brane action \eqref{M5action1}, as anticipated in the discussion before (\ref{bparameter}).

Second, we have to set the parameter which appear in the $\theta$-functions in \eqref{W_M5} to the values $\alpha_M = 0 = \beta^N$, thereby fixing the correct expression for the M-theory partition function. A non-trivial value for the half-integer vector  $\alpha_M$ would correspond to a shift in the Freed-Witten quantisation condition on the instanton flux $\cF_{E}$. However, this does not occur for $O(1)$ instantons, as in this case $c_1(K_{D_E}) \in H^2_+(D_E)$. In addition there appears to be no room for a non-vanishing vector $\beta$.
As discussed the procedure to correctly identify $\alpha_M, \beta^N$ purely within M-theory is non-trivial and quite involved in explicit setups \cite{Witten:1996hc}.
Our identification $\alpha = 0 = \beta$ is in agreement with previous analyses of an M5-brane wrapped on a flat torus. In this situation the result has been deduced by requiring modular invariance of the partition function \cite{Gustavsson:2000kr}, by comparison with D4-brane results \cite{Gustavsson:2011ur} or by a direct computation using a Hamiltonian formulation \cite{Dolan:1998qk}. It is also in agreement with the study of NS5-instantons performed in \cite{Dijkgraaf:2002ac}. In fact, it has been pointed out in \cite{Gustavsson:2000kr} that in the case where the worldvolume of the M5-brane can be written as $T^2\times M_4$ with a simply-connected manifold $M_4$, the requirement of modular invariance suffices to determine the partition function uniquely. However, our result is applicable to a much larger class of problems including degenerations of the torus fiber. Recall that the existence of such degeneration loci in the instanton worldvolume are necessary to be able to interpret the instanton as being of $O(1)$ type. In our case the geometry of the instanton is unconstrained apart from the requirement that the divisor $D_M$ be a vertical divisor\footnote{Relaxing the assumptions of rigidity and absence of Wilson lines would lead to more complicated notation as the classical partition sums would run only over suitable subspaces of $H^3(D_M)$, but would not change the qualitative results. See the discussion at the end of the present subsection for further details.}. Note that the existence of a Type IIB limit to compare with only implies certain topological constraints on the discriminant locus of the elliptic fibration \cite{Krause:2012yh}. As far as the structure of the partition function (\ref{W_M5}) is concerned, this is not a restriction.

The relationship between the $C_-^m$ and $G^m$ can also be obtained directly by examining the expansion of $C_3$,
\bea
\iota^* C_3 \supset c^a \iota^* \alpha_a + b_b \iota^* \beta^b & =&  M_a^m (c^a - \tau \delta^{ab}b_b) E_m^- + M_a^m (c^a - \bar{\tau} \delta^{ab}b_b)E_m^+ \nonumber \\
\label{pullback_C3_and_Cm}
& = & M_a^m G^a E_m^- + M_a^m \bar{G}^a E_m^+,
\eea
confirming the identification $C_-^m \equiv G^m$. 
Finally let us also check that the classical prefactor of the M5 action also matches precisely with the corresponding Type IIB expression. Expanding $[D_M] = C^\alpha_E \,  \omega_\alpha$ we get with the definitions of appendix \ref{democratic_reduction}
\be
\label{M5_action_term1}
-2\pi \int_{M5} ( \frac{1}{3!} J^3 + i C_6) = -\pi C^\alpha_E \left( 2\cV_\alpha + i \tilde{c}_\alpha \right) = -\pi C^\alpha_E \left( 2\cV_\alpha + i ( c_\alpha - \frac12 \cK_{\alpha ac}\delta^{cb} c^a b_b) \right),
\ee
which in the F-theory limit reduces precisely to $-\pi \left( \frac12 C^\alpha_E \cK_{\alpha \beta\gamma}v_B^\beta v_B^\gamma + iC^\alpha_E(c_\alpha - \frac12 \cK_{\alpha ab}c^a b^b)\right)$. This reproduces the moduli dependence of the Type IIB E3-instanton action using the natural identification of $c_\alpha,\ c^a,\ b^a$ whith their IIB counterparts. Note that the shift between $\tilde{c}_\alpha$ and $c_\alpha$ determined in appendix \ref{democratic_reduction} is crucial in obtaining this exact match.

Let us now consider the effect of relaxing the rigidity condition on the instanton. Recall that in this case $H^{3}(D_M)$ splits into the subspaces $H^{(3,0)}(D_M) \oplus H^{(2,1)}_{\rm hor}(D_M)$ and $H^{(2,1)}_{\rm ver}(D_M)$ (plus conjugates). These subspaces are the uplifts of $H^{(2,0)}_-(D_E)$ and of $H^{(1,1)}_-(D_E)$, respectively. The F-term supersymmetry condition restricts the flux on the E3-instanton to take values in $H^{(1,1)}_-(D_E)$. We expect that there exists a corresponding F-term condition on the M-theory side that restricts the supersymmetric flux on the M5-instanton to $H^{(2,1)}_{\rm ver}(D_M)$. In this case also $k_M$ appearing in the M5-instanton partition function (\ref{W_M5}) is a vector of dimension $\dim H^{2,1}_{\rm ver}(D_M)$, and the identification in (\ref{identifications_for_partition_function}) is unchanged.

The F-term condition for the E3-instanton can be obtained from the superpotential term
\beq 
\label{FtermE3}
\int_{D_E} i_v \Omega_3 \wedge \cF_{E},
\eeq
where $i_v \Omega_3$ denotes the contraction of a normal vector field to $D_E$ with the holomorphic (3,0)-form $\Omega_3$. 
(\ref{FtermE3}) allows for an obvious uplift of the form
\beq 
\label{FtermM5}
\int_{D_M} i_v \Omega_4 \wedge \cal H,
\eeq
which suffices to rule out unwanted contributions to $\cH$ in $H^{(3,0)}(D_M)$. However it is less clear how contributions from $H^{(2,1)}_{\rm hor}(D_M)$ are ruled out. It would be interesting to investigate whether the mechanism is connected to the fact that in Calabi-Yau fourfolds the variations of $\Omega_4$ do not span $H^{2,2}(Y_4)$ or whether further contributions to \eqref{FtermM5} must be included, but this is beyond the scope of this work.

In the above we have always assumed that $h^1_-(D_E \cap D_A)=0$. If one relaxes this assumption there must be additional contributions to \eqref{FtermM5} which rule out this additional set of 3-form fluxes from appearing in the partition function, in order to have a match with Type IIB where the fluxes are in $H^{(1,1)}_-(D_E)$. The same statement applies if we consider relaxing the assumption regarding absence of Wilson lines. However a detailed study of how these constraints arise is beyond the scope of this paper and would be an interesting subject for further study.

%%%%%%%%%%%%%%%%%%%%%%%%%%%%%%%%%%%%%%%%%%%%%%%%%%%%%%%%%%%%%%%%%%%%%%%%%%%%%%%%%%%%%%%%%%%%%%%%%%%%%%%%%%%%%%%%%%%%%%%%%%%%%
\section{Instanton selection rules in backgrounds with $G_4$-flux} \label{sec:Inst_plus_flux}
%%%%%%%%%%%%%%%%%%%%%%%%%%%%%%%%%%%%%%%%%%%%%%%%%%%%%%%%%%%%%%%%%%%%%%%%%%%%%%%%%%%%%%%%%%%%%%%%%%%%%%%%%%%%%%%%%%%%%%%%%%%%%

In this section we discuss the interplay between the M5-instanton and the 7-branes of the compactification due to non-trivial intersections of $D_M$ with the discriminant locus. In Type II orientifolds the effect of brane-instanton intersections is well-known \cite{Blumenhagen:2006xt,Ibanez:2006da, Florea:2006si}: The open string sector gives rise to zero modes localised on the intersection curve which are charged under the brane gauge group. In the presence of non-vanishing \emph{gauge flux} on the 7-branes, the spectrum of these modes can be chiral, resulting in a net charge of the instanton under the 7-brane gauge group. In \cite{Grimm:2011dj} it was pointed out that also the \emph{instanton flux} can lead to a chiral zero mode spectrum, and that the effect of gauge and instanton flux can cancel each other.
These two effects have rather different analogues for M5-instantons:
In F/M-theory 7-brane gauge flux is described by $G_4$-flux, whereas we have seen that instanton flux is described by 3-form flux. The effect of $G_4$-flux on the instanton partition function is the subject of section \ref{sec:M2_zero_modes}. As we will see and has already been discussed from various perspectives in \cite{Donagi:2010pd,Marsano:2011nn}, absence of chiral zero modes due to gauge flux requires that the pullback of the $G_4$-flux onto the M5-instanton vanishes, $\iota^* G_4=0$. In section \ref{sec:selection_rules_Ftheory} we analyse this constraint in detail in the framework of \cite{Krause:2011xj,Krause:2012yh}. In section \ref{sec:selection_rules_IIB} we compare the flux induced instanton charges in the languages of F/M-theory and Type IIB. This includes a discussion of massive $U(1)$s and also some speculative remarks about zero modes induced by \emph{instanton flux} from the M5-brane perspective.

%%%%%%%%%%%%%%%%%%%%%%%%%%%%%%%%%%%%%%%%%%%%%%%%%%%%%%%%%%%%%%%%%%%%%%%%%%%%%%%%%%%%%%%%%%%%%%%%%%%%%%%%%%%%%%%%%%%%%%%%%%%
\subsection{$G_4$-flux and zero modes from M2-branes}
\label{sec:M2_zero_modes}
%%%%%%%%%%%%%%%%%%%%%%%%%%%%%%%%%%%%%%%%%%%%%%%%%%%%%%%%%%%%%%%%%%%%%%%%%%%%%%%%%%%%%%%%%%%%%%%%%%%%%%%%%%%%%%%%%%%%%%%%%%

$G_4$-fluxes form a crucial ingredient of phenomenologically interesting F-theory backgrounds as they are necessary to generate chiral matter spectra. As such the effects and explicit construction of consistent $G_4$ gauge flux as suitable elements of $H^4(Y_4)$ have received considerable recent attention in the literature, see e.g. \cite{Braun:2011zm,Marsano:2011hv,Krause:2011xj,Grimm:2011fx,Krause:2012yh} for a selection of recent papers on the subject.

To appreciate how the inclusion of $G_4$ gauge flux modifies the M5-partition function we note that the action $S_{\cal B}$  in (\ref{M5action1}) implicitly depends on the pullback $\iota^*G_4$.
Integration by parts of the last term  gives $2 \pi i \int   {\cal H} \wedge \iota^* C_3 = - 2 \pi i  \int {\cal B} \wedge  \iota^* dC_3$. As always this should be read as a piece depending on the fluctuations of $dC_3$ and a piece involving the background flux $G_4 = \langle dC_3 \rangle$. In the sequel we make this distinction explicit by writing
\bea
S_{{\cal B}, G_4} = S_{\cal B} -  2 \pi i   \int {\cal B} \wedge \iota^*G_4
\eea
with the understanding that $S_{\cal B}$ contains the fluctuation term upon integration by parts. It is this piece that we have considered so far. Note that the additional term is present only if $\iota^* G_4 \neq 0$ in the cohomology of $D_M$, which we assume to be the case in the remainder of this subsection. 

Let us first point out one important consequence of $\iota^* G_4 \neq 0$: The coupling to $G_4$ breaks the group of large gauge transformations of ${\cal B}$ (without shifting $C_3$ or $C_6$) from $H^2(D_M, \mathbb R)$ to a discrete subgroup, say,  $H^2(D_M, \mathbb Z)$ (or an analogous one depending on the explicit quantisation of $\iota^*G_4$). This implies that $H^2(D_M, \mathbb R)/ H^2(D_M, \mathbb Z)$ must now be integrated over in the path integral over ${\cal B}$.
The naive superpotential  for  $\iota^* G_4 \neq 0$ then derives from the integral
\be
\label{S_M5-G}
 \int{\mathcal D}{\mathcal B}  \,  e^{- [ 2\pi (\mathrm{Vol}_{D_M} + i \int C_6) + S_{\cB, G_4}]} \equiv  \int{\mathcal D}{\mathcal B} \, e^{- S_{M5, G_4}}.
\ee

This, however, is incomplete for the following two reasons:
First, as pointed out in \cite{Donagi:2010pd}, $S_{{\cal B}, G_4}$ contains a $G_4$-dependent tadpole for ${\cal B}$. As a consequence, the path integral $\int{\mathcal D}{\mathcal B} \, e^{- 2 \pi i \, {\cal B} \wedge \iota^*G_4}$ is proportional to a delta function $\delta(\iota^*G_4)$, which vanishes for non-trivial pullback of  $G_4$.  

We offer the following interpretation of this observation in the language of E3-brane instantons in the Type IIB limit:
In the presence of non-trivial gauge flux on the intersection curves of the instanton divisor and the 7-branes extra zero modes  $\lambda_a, \tilde \lambda_b$ appear, which are localised on these intersection curves and are charged under the 7-brane gauge group \cite{Blumenhagen:2006xt,Ibanez:2006da,Florea:2006si}.
Thus the partition function must include the integral over those zero modes $\lambda_a, \tilde \lambda_b$, and the direct analogue of the expression (\ref{S_M5-G}) is
\bea \label{lambda-SE}
\int {\mathcal  D} \lambda_a {\mathcal D} \tilde \lambda_b  \sum_{\cF_E} e^{- S_E [\cF_E]}.
\eea

In particular we can formally identify $\int {\mathcal  D} \lambda_a {\mathcal D} \tilde \lambda_b$ with the new integral $\int{\mathcal D}{\mathcal B}  e^{2 \pi i   \int {\cal B} \wedge \iota^*G_4}$ taken only over  $H^2(D_M, \mathbb R)/ H^2(D_M, \mathbb Z)$.
As we will see in section \ref{sec:selection_rules_Ftheory} this is further supported by the fact that the charge of this piece under massless $U(1)$s, if present, cancels the charge of $e^{- [ 2\pi (\mathrm{Vol}_{D_M} + i \int C_6) + S_{\cB}]}$, which now also exhibits a non-trivial shift due to the transformation of $C_6$.
Without further modifications the quantity (\ref {lambda-SE}) vanishes as the Grassmann integral is not saturated.

Second, and independently of the previous argument, there is an intrinsic reason why the expression for the superpotential must be modified both in M-theory and in the Type II limit. For the M5-brane the equations of motion for $\cB$ from \eqref{M5action1}, 
\be
d \ast \cH \propto \iota^* G_4,
\ee
call for additional sources cancelling $[\iota^* G_4]$ for consistency. In fact, as discussed in \cite{Marsano:2011nn}, the M-theory version of the Freed-Witten anomaly \cite{Freed:1999vc} directly requires the inclusion of such sources in the form of M2-branes ending on the instanton. 
These  M2-branes cancel the flux-induced source term. The world-volumes of the M2-branes in question are given by 3-chains of the form $\Gamma_i = I \times \gamma_i$, where the $\gamma_i$ are 2-cycles in $D_M$ satisfying \cite{Marsano:2011nn}
\be
\label{M2_cycles}
\sum_i \gamma_i = \mathrm{PD}_{D_M}[\iota^* G_4 ]
\ee
and $I$ is a semi-infinite time interval such that $\partial \Gamma_i = \{t_0\} \times \gamma_i$ in the M5-instanton worldvolume. $\mathrm{PD}_{D_M}$ denotes the Poincar\'e dual taken on the divisor $D_M$.

The appearance of these M2-brane states can also be seen directly by considering the solution for $\cH$ in a background with $G_4$. This flux configuration will be sourced along $\mathrm{PD}_{D_M}[\iota^* G_4 ]$. In a supersymmetric configuration this field configuration is accompanied by a deformation of the M5-brane into the normal direction. This deformation can be identified with an M2-brane ending on the brane at $\mathrm{PD}_{D_M}[\iota^* G_4 ]$ \cite{Howe:1997ue}. In particular this picture suggests another equivalent way of thinking about the zero modes of the M5-instanton, namely as the zero modes describing the deformations of the solitonic $\cH$-field solution sourced by the $G_4$-flux.

The upshot of the discussion above is that when calculating any correlation functions in the M5-instanton background we must include the operators describing these M2-brane states, which are proportional to the exponential of the action of the M2-brane \cite{Marsano:2011nn}. This leads to \cite{Bergshoeff:1987qx}
\be
\label{M2_vertex_factors}
V_{M2,i} \propto \exp \left[ 2\pi i \left( \frac12 \int_{\Gamma_i} C_3  + \int_{\gamma_i} \cB \right) \right].
\ee
These states therefore contribute to any instanton-generated coupling in the presence of non-trivial $G_4$-flux on the instanton. Such a coupling then takes the form
\be
\label{superpot_with_M2}
W_{M_5} \propto \int \cD \cB \ \prod_i V_{M2,i} \ e^{-S_{M5}}.
\ee

We suggest an interpretation of these vertex operators by comparing again the situation for M5-branes with the Type IIB limit: The effective action of the E3-brane instanton contains interaction terms between the charged zero modes $\lambda_a, \tilde \lambda_b$ and charged open string modes $\Phi_{ab}$ in the sector of open strings stretching between 7-brane stacks $a$ and $b$.  These operators make the saturation of the Grassmann integral possible.
A superpotential term generated by such an instanton then schematically takes the form \cite{Blumenhagen:2006xt,Ibanez:2006da,Florea:2006si}
\be
\label{superpot_charged_modes}
W_{E3} \propto \int \cD \lambda_a \cD \tilde{\lambda}_b \ e^{-S_{E3} + \int \lambda_a \Phi_{ab} \tilde{\lambda}_b}.
\ee
A non-vanishing result therefore hinges upon the insertion of the vertex operator 
\bea
e^{\int \lambda_a \Phi_{ab} \tilde{\lambda}_b}
\eea
in the path integral (\ref{lambda-SE}).
The M2-brane states are therefore the M-theory analogue of the charged open string states appearing in \eqref{superpot_charged_modes}
\be
\label{M2_openstrings}
V_{M2,i} \cong e^{\int \lambda_a \Phi_{ab} \tilde{\lambda}_b}.
\ee
Note that in both pictures the full vertex operator is gauge invariant in particular with respect to extra $U(1)$ symmetries, if present, as discussed in the next section. 

%%%%%%%%%%%%%%%%%%%%%%%%%%%%%%%%%%%%%%%%%%%%%%%%%%%%%%%%%%%%%%%%%%%%%%%%%%%%%%%%%%%%%%%%%%%%%%%%%%%%%%%%%%%%%%%%%%%%%%%%%
\subsection{Selection rules for the absence of chiral charged zero modes}
\label{sec:selection_rules_Ftheory}
%%%%%%%%%%%%%%%%%%%%%%%%%%%%%%%%%%%%%%%%%%%%%%%%%%%%%%%%%%%%%%%%%%%%%%%%%%%%%%%%%%%%%%%%%%%%%%%%%%%%%%%%%%%%%%%%%%%%%

From the discussion in the previous subsection it is clear that a necessary condition for an M5-instanton to contribute to a superpotential coupling \emph{without} involving any M2-brane vertex operators is
\be
\label{sel_rule_pullback_G4}
\iota^* G_4 = 0.
\ee
While this well-known form of the Freed-Witten condition is conceptually simple, it can nevertheless be non-trivial to evaluate in practice as it requires checking that the integral of $G_4$ over every 4-cycle in $D_M$ vanishes.\footnote{Note that even if (\ref{sel_rule_pullback_G4}) is satisfied, the inclusion of extra vertex operators, albeit with zero net charge under the 7-brane gauge groups, may be required, as is familiar from Type II instantons with vector-like pairs of charged zero modes. We will only analyze the above necessary condition in the sequel.}

\subsubsection{General picture}

To analyse the selection rule in detail let us first recall the explicit description of $G_4$ gauge fluxes  F/M-theory.
We are working in the framework of an F/M-theory compactification on an elliptically fibered Calabi-Yau fourfold $Y_4$ whose singularity structure is responsible for the appearance of a gauge group $G$ along the 7-branes. This may include also Abelian factors of $U(1)_A$ gauge groups. A proper definition of the low-energy effective theory is possible after resolving the singularities of $Y_4$ and by working on the resolved fourfold $\hat Y_4$. The singularities associated with the non-Abelian piece $G'$ of the total gauge group
are resolved by introducing a set of  $\rm rk{(G')}$ resolution divisors $E_i$. One way to realise extra $U(1)$s associated with non-Cartan generators\footnote{These will be referred to as "non-Cartan $U(1)$s" in the sequel.} in Calabi-Yau fourfolds has been described in \cite{Grimm:2010ez}; this leads in general to a set of further resolution divisors $S_A$.

Gauge flux on $\hat Y_4$ is encoded in a class $G_4 \in H^{(2,2)}(\hat Y_4)$ vertical to the pullback of any two base divisors and to the section of the elliptic fibration.
The cohomology $H^{(2,2)}(\hat Y_4)$ splits into the primary vertical subspace $H^{(2,2)}_{\rm ver}(\hat Y_4)$  and  the primary horizontal one $H^{(2,2)}_{\rm hor}(\hat Y_4)$, defined as the complement of the first \cite{Greene:1993vm,Mayr:1996sh,Grimm:2009ef,Intriligator:2012ue}. By definition, $H^{(2,2)}_{\rm ver}(\hat Y_4)$ is spanned by 4-forms that factorise into two 2-forms.
A special type of gauge fluxes in $H^{(2,2)}_{\rm ver}(\hat Y_4)$ is of the form
\be
\label{massless Flux}
G_4 = -{\cF}^{\rm b}_{A} \wedge \tw_{A}, \qquad {\cF}^{\rm b}_{A} \in H^{(1,1)}(B_3)
\ee
with  $\tw_{A}$ an element of $H^{1,1}(\hat Y_4)$ which is not in $H^{1,1}(B_3)$ and not given by the fiber class. Such $G_4$ is the flux associated with a $U(1)_A$ gauge symmetry whose potential arises from the expansion
\bea \label{C3A}
C_3 = A_A \wedge  \tw_{A} + \ldots.
\eea
This is true for Cartan $U(1)$s of a non-Abelian gauge group or for non-Cartan $U(1)$s alike. For Cartan fluxes, the 2-form $\tw_{A}$ is the 2-form dual to a combination of resolution divisors $E_i$ introduced in the blow-up of the non-Abelian singularity over a given divisor. 
For non-Cartan $U(1)$s the 2-form $\tw_A$ is related to the resolution divisors $S_A$ \cite{Grimm:2010ez}, where the exact form of $\tw_A$ determines the precise normalisation of the $U(1)_A$ charges. Associated $U(1)_A$ fluxes have been constructed in \cite{Braun:2011zm,Krause:2011xj,Grimm:2011fx}.

Flux of the type $G_4 = -{\cF}^{\rm b}_{A} \, \wedge \tw_{A}$ represents only a special class of gauge fluxes. 
More generic $G_4$ in $H^{(2,2)}_{\rm ver}(\hat Y_4)$ can also be the effectively vertical  sum of products of 2-forms without factorising into a 2-form in $H^{1,1}(B_3)$ and a 2-form in its complement in $H^{1,1}(\hat Y_4)$. In \cite{Marsano:2011hv} such gauge fluxes were constructed on resolved fourfolds and related to the  spectral cover approach. These fluxes are not associated with a massless $U(1)$ in F-theory. 
As shown in \cite{Krause:2012yh}, in the Type IIB limit such flux can be identified with the linear combination of diagonal $U(1)$s which is massive by the geometric St\"uckelberg mechanism, where in addition one must impose the D5-tadpole constraint.\footnote{ In  \cite{Grimm:2011tb} it was conjectured that the uplift of such fluxes should be described by a set of non-harmonic fluxes subject to an analogue of the  D5-tadpole constraint. This is consistent with the recent observations of \cite{Krause:2012yh} because the D5-tadpole constraint of \cite{Grimm:2011tb} has the potential to effectively eliminate the non-harmonic parts.}
 
Finally $G_4$ can be an element of $H^{(2,2)}_{\rm hor}(\hat Y_4)$. In particular, the Higgsing of a $U(1)$ symmetry via brane recombination results in a geometry that does not exhibit the correponding 2-form $\tw_{A}$ any longer. As a result of this deformation 4-forms of the type $D \wedge \tw_A$ become non-factorisable elements in $H^{(2,2)}_{\rm hor}$. 
Examples of gauge fluxes described by these forms have been constructed in  \cite{Braun:2011zm} (see also \cite{Krause:2011xj}).

We now come back to an analysis of the constraint (\ref{sel_rule_pullback_G4}).
To evaluate it we consider integrals of the form
\be
\label{sel_rule_gen_surface}
\int_{C_{(4)}} \iota^* G_4 \qquad \text{ with } C_{(4)} \in H_4 (D_M).
\ee
Since $\iota^* G_4$ is obtained by pullback from the ambient Calabi-Yau $\hat Y_4$, this integral can only give a non-vanishing result if $C_{(4)}$ is non-trivial in $H_4 (\hat Y_4)$. By construction the fluxes are orthogonal to surfaces either lying entirely inside the base or wrapping the full elliptic fiber.

As a result only two types of surfaces can lead to a non-zero integral \eqref{sel_rule_gen_surface}. 
The first type of surfaces exists in the presence of massless $U(1)$ factors (including the Cartan $U(1)$s) and the associated selection rule is related to the net $U(1)$ charge of the instanton. 
Since with each massless $U(1)_A$ symmetry comes a 2-form $\tw_A \in H^{1,1}(\hat Y_4)$ as in (\ref{C3A}), one can construct the algebraic surface with dual 4-form 
\be \label{C4A-gen}
[C_{(4)}^A] = -  D_M \wedge \tw_A  \in H^{(2,2)}_{\rm vert}(\hat Y_4).
\ee
The surfaces $[C_{(4)}^i] = -  D_M \wedge E_i$ are associated with the Cartan $U(1)$s within the non-Abelian group $G'$, but also the non-Cartan $U(1)$s give rise to corresponding surfaces.
Indeed the integral
\bea \label{C4Aconstr}
\int_{C_{(4)}^A} \iota^* G_4  = -  \int_{\hat Y_4} D_M \wedge \tw_A \wedge G_4
\eea
is in general non-vanishing. For fluxes which have been chosen such as to leave the non-Abelian gauge group $G'$ unbroken, this integral vanishes by construction for the Cartan surfaces $[C_{(4)}^i]$. Nevertheless a non-trivial constraint remains in the presence of extra non-Cartan $U(1)$s.
Indeed the expression (\ref{C4Aconstr}) has a very intuitive interpretation from the point of view of the three-dimensional low-energy effective action obtained by compactification of M-theory because $G_4$-flux induces a gauging of the shift symmetry of the fields $\tilde{c}_\alpha$ appearing in an expansion of $C_6$. As discussed in Appendix \ref{democratic_reduction}, the gauging of the fields $\tilde{c}_\alpha$ under a gauge transformation of a canonically normalised $U(1)_A$ is given by
\be
\label{gauge_trf_A}
A_A\rightarrow A_A + d\Lambda_A \ \Rightarrow \ \delta_{\Lambda_A} \tilde{c}_\alpha = - \Theta_{\alpha  A} \Lambda_A.
\ee 
The charge vector $\Theta_{\alpha A}$ is defined by
\be 
\Theta_{\alpha A} = 2 \int_{\hat Y_4} \omega_\alpha \wedge \tw_A \wedge G_4.
\ee
The induced shift of the term $\int C_6$ in the instanton action is then simply
\be \label{U1chargea}
\delta_{\Lambda^A} \int C_6 = -\frac12 C^\alpha_E \Theta_{\alpha A} \Lambda^A = - \int_{\hat Y_4} D_M \wedge \tw_A \wedge G_4,
\ee
which is none other than \eqref{C4Aconstr}. 
From the Type II perspective we are well familiar with the fact that a net transformation of the instanton action under a given $U(1)_A$ symmetry requires the introduction of charged operators ${\cal O}$ in the non-perturbative superpotential such that ${\cal O} \, e^{-S}$ is gauge invariant. As discussed in section (\ref{sec:M2_zero_modes}) such a mechanism is also required from a microscopic perspective. We see that the microscopic and the supergravity constraint match, as they must.

The second type of surfaces are present also in the absence of a massless $U(1)$. These surfaces lie within $D_M$ even though they cannot be described as algebraic surfaces for generic choices of complex structure moduli. Similar phenomena have recently played important roles in the context of $G_4$ gauge fluxes in \cite{Braun:2011zm,Collinucci:2012as}.
Nonetheless they give a non-zero contribution in integrals of the form (\ref{sel_rule_gen_surface}).
One type of such surfaces is related to the matter surfaces of the 7-brane sector; another class can be related to the algebraic surfaces (\ref{C4A-gen}) upon brane recombination, or Higgsing of the $U(1)_A$, and can be described with the methods of \cite{Braun:2011zm}. In both cases the selection rule has no interpretation as the charge of a \emph{massless} $U(1)$ in M-theory.
Note that the number of independent selection rules will in general be much smaller than the number of surfaces that can be constructed in this way.
 
 This second type of surfaces is more complicated because the dual 4-forms are no longer factorisable. For ease of presentation we therefore refrain from describing them in full generality but rather illustrate them in a concrete set of models, furnished by a global Tate model with gauge group $SU(5) \times U(1)_X$. 
 The generalisation to more general groups (e.g. $SU(N)$ instead of $SU(5)$) and to several $U(1)$ factors will be clear.

\subsubsection{Example: $SU(5) \times U(1)_X$}
\label{sec_su5xu1}

Our treatment of the $SU(5) \times U(1)_X$ Tate model is based on the analysis of refs. \cite{Krause:2011xj,Krause:2012yh}, whose notation we briefly summarise.
Consider a singular Calabi-Yau fourfold $Y_4$ given as a Tate model over a general base $B_3$,
\bea
 P_T : \{y^2 + a_1 x y z + a_3 y z^3 = x^3 + a_2 x^2 z^2 + a_4 x z^4 + a_6 z^6\}.
\eea
The locus $P_T$ defines a hypersurface in an ambient fivefold, which in turn is given by fibering the weighted projective space $\mathbb P_{2,3,1}$ with homogeneous coordinates $x,y,z$  over $B_3$.  The Tate polynomials $a_i$ are sections of $\bar {\cal K}^i$, where $\bar {\cal K}$ is the anti-canonical bundle of $B_3$.
They are chosen in such a way as to produce 
an $SU(5)$ singularity in the fiber over the base divisor ${\cal W}: w=0$. This leads to \cite{Bershadsky:1996nh,Andreas:2009uf}
\bea \label{Tatepolys}
a_1 \,\, \,  {\rm generic}, \qquad a_2 = a_{2,1} \, w, \qquad a_3 = a_{3,2} \, w^2, \qquad a_4 = a_{4,3} \, w^3, \qquad a_6 =a_{6,5} \, w^5.
\eea
Calabi-Yau fourfolds for F-theory compactifications of this type have been analysed intensively in the context of F-theory model building in the recent literature \cite{Marsano:2009ym,Marsano:2009gv,Blumenhagen:2009yv,Grimm:2009yu,Chen:2010ts,Knapp:2011wk}.
Furthermore we take $a_6$ subject to the extra constraint $a_{6,5} =0$ such as to produce an additional $U(1)_X$ gauge group \cite{Grimm:2010ez}.
After resolution of the singularities associated with the gauge groups $SU(5) \times U(1)_X$, the resolved Calabi-Yau fourfold $\hat Y_4$ is given by the proper transform of the Tate polynomial in a toric ambient fivefold $X_5$. In addition to the usual coordinates $x,y,z$ of the smooth Tate model and the coordinates of the base $B_3$, the homogeneous coordinates of this fivefold include the coordinates $e_i,\ i=1,\ldots ,4$ (as blow-up coordinates of the $SU(5)$ singularity) and $s$ as a consequence of the Abelian gauge group, together with a number of scaling relations. Associated with these resolution coordinates are the respective divisor classes $E_1,...,E_4,S$. The first explicit resolution of these singularities in Calabi-Yau fourfolds has been provided in the toric examples of \cite{Blumenhagen:2009yv,Grimm:2009yu} (see also \cite{Collinucci:2010gz}), followed by an in-depth analysis of the resolution in particular of higher codimensions singularities in \cite{Esole:2011sm,Marsano:2011hv, Krause:2011xj,Grimm:2011fx}. The resolved Calabi-Yau fourfold $\hat Y_4$ is then described by the locus
\begin{equation}\label{tate_proper_transform}
 \bal
P_T: \{  y^2\,s\,e_3\,e_4 &+ a_1\,x\,y\,z\,s + a_{3,2}\,y\,z^3\,e_0^2\,e_1\,e_4 \\
                   &= x^3\,s^2\,e_1\,e_2^2\,e_3 + a_{2,1}\,x^2\,z^2\,s\,e_0\,e_1\,e_2 + a_{4,3}\,x\,z^4\,e_0^3\,e_1^2\,e_2\,e_4\}
 \eal
\end{equation}
within $X_5$, where $e_0$ is the coordinate associated with the proper transform of the divisor ${\cal W}$.
Finally the extra $U(1)_X$ factor is due to the existence of the 2-form
\bea \label{twX}
 \tw_X = S - Z - \bar{\cK} + \frac{1}{5}(2,4,6,3)_i E_i,
 \eea
which is related to the resolution divisor $S$ arising in the $U(1)$-restricted Tate model. The specific linear combination $\tw_X$ is chosen such that the $U(1)_X$ generator is orthogonal to the Cartan generators of $SU(5)$ \cite{Krause:2011xj}.

There exist exactly two types of generic (i.e. base-independent) $G_4$-fluxes in $H^{(2,2)}_{\rm ver}(Y_4)$ which do not break the $SU(5)$ gauge group. In terms of the above divisor classes they are given by \cite{Marsano:2011hv, Krause:2011xj,Krause:2012yh,Grimm:2011fx}
\bea
\label{G_4_X}
G^X_4  &=&  - \cF^{\rm b} \wedge \tw_X ,\\
G^\lambda_4 &=& \lambda (E_2\wedge E_4 + \frac{1}{5} (2,-1,1,-2)_i E_i \wedge \bar{\cK} ).
\label{G_4_lambda}
\eea
It is understood that the parameter $\lambda$ and $\cF^{\rm b} \in H^{1,1}(B_3)$ are chosen in such a way that the fluxes obey the Freed-Witten quantisation condition
\be
G_4 + \frac12 c_2(\hat Y_4) \in H^4 (\hat Y_4, \mathbb{Z}).
\ee
This constraint has been analysed in detail in \cite{Collinucci:2010gz,Collinucci:2012as}.
The fluxes (\ref{G_4_X}) and (\ref{G_4_lambda}) are what we will focus on in the sequel.

Following the general discussion of the previous subsection, associated with the $U(1)_X$ gauge factor is the algebraic surface with dual 4-form 
\be
[C_{(4)}^X] = -  D_M \wedge \tw_X.
\ee
General integrals of the type $- \int_{\hat Y_4} D \wedge \tw_X \wedge G_4$ have first been computed in \cite{Krause:2012yh} for the fluxes in (\ref{G_4_X}) and (\ref{G_4_lambda}). 
Using these results one finds
\bea
\label{sel_rule_G^X_over_C^X}
5 \int_{C_{(4)}^X} \iota^* G_4^X  &=& - 10 \int_{B_3} (\bar{\cK} - \frac{3}{5} {\cal W}^{\rm b} )\wedge D^{\rm b}_M \wedge \cF^{\rm b} , \\
5 \int_{C_{(4)}^X} \iota^* G_4^\lambda &=& -  {\lambda} \int_{B_3} \bar{\cK} \wedge {\cal W}^{\rm b} \wedge D_M^{\rm b} \label{sel_rule_G^lambda_over_C^X}.
\eea
According to the discussion around (\ref{U1chargea}) these expressions are the $U(1)_X$ charges of the instanton in presence of the corresponding fluxes, where the explicit factor of $5$ has been included to account for the normalisation of the $U(1)_X$ boson associated with $\tw_X$ as defined in (\ref{twX}).

The second set of surfaces over which we can integrate $\iota^* G_4$ is related to the matter surfaces. Matter surfaces consist of certain combinations of $\mathbb{P}^1$s fibered over curves in the base where the singularity type is enhanced. They have recently been analysed in detail in the context of elliptic fourfolds  in \cite{Esole:2011sm,Marsano:2011hv, Krause:2011xj,Grimm:2011fx}. We follow the approach and notation of  \cite{Krause:2011xj}.
In the $SU(5) \times U(1)_X$ setup under consideration, there are three types of such matter surfaces, corresponding to the loci of singularity enhancement to $SO(10)$ and $SU(6)$ on the GUT brane as well as the curve of self-intersection of the $I_1$-locus with an enhancement to $SU(2)$. 
We exemplify the constraints arising from these surfaces by analysing the surfaces $C_{10}^k$ fibered over the curve of $SO(10)$ enhancement $\cC_{10} = {\cal W}^{\rm b} \cap \{ a_1 = 0 \} \subset B_3$. 
Each resolution divisor $E_i$ is $\mathbb P^1$ fibered over $\cal W^{\rm b}$. Some of these 
 $\mathbb P^1$s in the fiber split into two or several $\mathbb P^1$s over $\cC_{10}$. The  $\mathbb P^1$s in the fiber over $\cC_{10}$ group into certain linear combinations of curves which can be identified with the weights of the ${\bf 10}$-representation of $SO(10)$. By this one means that an M2-brane wrapping one of these linear combinations of $\mathbb P^1$s gives rise to a massless state corresponding to one of the ${\bf 10}$-components. The fibration of these curves over ${\cal C}_{10}$ is what one calls the matter surfaces $C_{10}^k, k=1, \ldots 10$.

As an example of one these ten surfaces, consider the complete intersection within the ambient fivefold $X_5$ given by
\bea \label{P13a}
{C}_{24}: \{e_2=0\} \cap \{e_4=0\} \cap \{ a_1=0  \}.
\eea
As discussed in \cite{Krause:2011xj}, $C_{24}$ describes a $\mathbb P^1$ fibered over the curve ${\cal C}_{10}: \{a_1 =0\} \cap {\cal W}^{\rm b}$ on $B_3$, as is
 manifest from the form of the resolved Tate constraint $P_T$ (\ref{tate_proper_transform}): If $e_2=e_4=a_1=0$, the Tate constraint is satisfied so that the locus lies inside $\hat Y_4$. Furthermore all $e_i$ are $\mathbb P^1$ fibered over ${\cal W}^{\rm b}$, hence the surface ${C}_{24}$ is itself fibered over the curve 
${\cal C}_{10}$ in the base.
Note that this surface can be written as a complete intersection of the ambient fivefold $X_5$, but not as a complete intersection on the Calabi-Yau $\hat Y_4$ itself.

After this brief review we finally turn to the instanton wrapping the divisor $D_M$. We will show under a specific technical assumption that if $D^b_M$ intersects the $SU(5)$ divisor ${\cal W}^b$ one can construct surfaces lying within $D_M$ that are analogous to the matter surfaces. However for generic choice of complex structure moduli these surfaces cannot be realised as algebraic surfaces. 
By smoothly deforming the complex structure moduli to the special locus where the matter surfaces become algebraic one can compute the integral (\ref{sel_rule_gen_surface}) of $\iota^* G_4$ over the matter surfaces within $D_M$ and deduce a selection rule for the instanton.

As an example we consider the locus
\bea
\label{c_24_dm}
{C}_{24}|_{D_M}: \{e_2=0\} \cap \{e_4=0\} \cap \{ D_M=0  \}
\eea 
in the ambient fivefold, which is the obvious candidate for a surface analogous to ${C}_{24}$ lying inside the instanton. Of course the following arguments immediately carry over if one starts with one of the other matter surfaces $C_{10}^k$ rather than ${C}_{24}$. If we assume for the moment that this surface can be taken to lie inside the instanton divisor $D_M \subset \hat Y_4$, we obtain a selection rule by integrating $\iota^* G_4$ over it. For the fluxes in $H^{(2,2)}_{\rm ver}(\hat Y_4)$ considered here this integral is a sum of terms, each of which can be written as a complete intersection of five divisors within the ambient fivefold $X_5$
\be \label{intC24DM}
\int_{{C}_{24}|_{D_M}} \iota^* G_4 = \int_{X_5}   E_2 \wedge E_4 \wedge D_M \wedge G_4.
\ee
The integral $\int_{{C}_{24}} \iota^* G_4$ has been performed in  \cite{Krause:2011xj} and expressed as the integral of a flux-dependent 2-form on $B_3$ over the locus ${\cal C}_{10}$. The result of (\ref{intC24DM}) is therefore the same integral over the curve $D^{\rm b}_M \cap {\cal W}^{\rm b}$ in $B_3$.
We thus conclude
\bea
 \int_{{C}_{24}|_{D_M}} \iota^* G^\lambda_4 =  \frac{\lambda}{5} \int_{B_3}  D_M^{\rm b} \wedge {\cal W}^{\rm b} \wedge (6 \bar{\cK} - 5 {\cal W}^{\rm b}), \quad    \int_{{C}_{24}|_{D_M}} \iota^* G^X_4 = \frac{1}{5} \int_{B_3}  D_M^{\rm b} \wedge {\cal W}^{\rm b} \wedge {\cal F}^{\rm b}.\nonumber \\
 \label{sel_matter_surface}
\eea
The value of the integral would be unchanged if we had taken one of the other matter surfaces $C_{10}^k$ instead of (\ref{P13a}).

Let us now justify our assumption that the algebraic cycle \eqref{c_24_dm} in $X_5$ could be used to obtain a selection rule. 
Note that while this surface manifestly lies inside $D_M \subset X_5$, for generic complex structure moduli it is not completely contained within $\{D_M = 0\} \cap \hat{Y}_4$. We will argue that there is however an associated non-algebraic surface within $\{D_M = 0\} \cap \hat{Y}_4$, over which we can integrate $G_4$ with the same result as if we had integrated over \eqref{c_24_dm}. Let us consider a special geometric setup for which there exist holomorphic polynomials $\alpha$, $\beta$ with associated effective divisor classes $[\alpha], [\beta]$ and an integer $n$ such that we can write
\be
\label{specialisation1}
(\tilde{a}_1)^{n+1} = \alpha D^{\rm b}_M + \beta {\cal W}^{\rm b},
\ee
with $\tilde{a}_1$ in the class $\bar{\cK}$. Now we can consider an auxiliary fourfold defined by $\tilde P_T$, where $\tilde P_T$ is given by \eqref{tate_proper_transform} with $a_1$ replaced by $\tilde a_1$. The condition \eqref{specialisation1} ensures that ${C}_{24}|_{D_M}$ is entirely contained within the deformed fourfold $\{ \tilde P_T \}$. As $\{ \tilde P_T \}$ is obtained from $\hat Y_4$ by a smooth deformation this shows that there is a non-holomorphic surface associated to ${C}_{24}|_{D_M}$ already contained in the original fourfold. In the course of this deformation the 4-forms dual to the surfaces picks up pieces in $H^{(3,1)}$ while keeping a piece in $H^{(2,2)}$. 
What counts is that the value of the topological integral (\ref{sel_matter_surface}) is unchanged under such a deformation, as $G_4$ is of type (2,2). Therefore, even though the surfaces $C_{10}^k$ are in general not realised as algebraic surfaces within $D_M$ for generic complex structure moduli, the integral $\int_{C_{10}^k} \iota^* G_4$ takes the value (\ref{sel_matter_surface}).

It remains to comment on how restrictive the assumption \eqref{specialisation1} is. Note that if the instanton and GUT divisor are chosen in such a way that
\bea \label{eff1}
[D^{\rm b}_M] \leq [a_1], \qquad \quad [{\cal W}^{\rm b}] \leq [a_1]
\eea
in the class-theoretic sense, then \eqref{specialisation1} can obviously be fulfilled with $n=1$. Now in most applications to model building one considers bases constructed as complete intersections in a toric ambient space and rigid instanton divisors $D_M^{\rm b}$. At least in this toric setting it is clear that rigidity implies that the polynomial $D_M^{\rm b}$ should be chosen to have as low a homogeneous degree as possible, so that \eqref{eff1} will often be fulfilled\footnote{In order to construct a consistent Weierstrass model also the class $[{\cal W}^{\rm b}]$ cannot be too large as the class $[a_{2,1}] = 2 \bar{\cK} - [{\cal W}^{\rm b}]$ must be effective.}.
To be slightly more general let us note that in many models $\bar{\cK}$ will be a big divisor, i.e. it will lie in the interior of the pseudo-effective cone. This is certainly the case for all base spaces that are obtained from Fano surfaces by blow-up. In fact, to the best of our knowledge, all examples of F-theory GUT models presented so far in the literature are of this form. In this case it is always possible to find an integer $n$ such that $[a_1^n] - [{\cal W}^{\rm b}]$ and $[a_1^n] - [D_M^{\rm b}]$ are effective. In other words, we can always find effective classes $\alpha$, $\beta$ and $\delta$ such that
\be
(a_1)^n = \alpha D^{\rm b}_M + \beta {\cal W}^{\rm b} + \delta.
\ee
Therefore the problem is reduced to the question of whether by a suitable deformation of the polynomials involved one can achieve $a_1^n - \delta = \tilde{a}_1^n$. While we expect this to be generically possible due to the freedom of choosing suitable representatives for $\alpha$, $\beta$, $D_M^{\rm b}$ and $\cW^{\rm b}$, it would be interesting to obtain a rigorous proof, and to see how this should be generalised in case $\bar {\cal K}$ is not a big divisor.

Let us summarise the upshot of the considerations presented above. Although in general the matter surfaces will not lie inside the instanton as holomorpic surfaces, we expect that one can smoothly deform the geometry in such a way that holomorphicity becomes manifest. This implies that the matter surfaces were already present inside the instanton in the undeformed geometry, albeit without a holomorphic representative. In the course of such deformations the Poincar\'e dual of the surface picks up pieces of Hodge type (1,3) and (3,1), but this does not change the value of the integral of the (2,2)-form $G_4$. Hence we can safely evaluate the integrals \eqref{sel_rule_gen_surface} in the geometry \eqref{specialisation1}. 
It would be interesting to understand how to generalise the proof to the general case.

Our presentation has focused on the surfaces fibered over the matter curve $\cC_{10}$ in the above, but similar arguments hold for the remaining matter surfaces.
All these integrals must vanish in order for the instanton to contribute to the superpotential without insertion of additional M2-Brane vertex factors. 
Note that the number of actual selection rules will in general be much smaller than the number of surfaces that can be constructed in this way, as matter surfaces from different enhancement curves are in general not linearly independent and the integrals of $G_4$ over all matter surfaces of a given enhancement curve are the same.
In the next subsection we will discuss the relation to the known selection rules in the Type IIB dual and we will confirm that the selection rule from $U(1)_X$ charge and from the $C_{10}$ surface generate the entire set of selection rules for models with an orientifold dual. This is intuitive because the $C_{10}$ surface is sensitive to charged zero modes between the instanton and the $SU(5)$ brane stack only, while the $U(1)_X$ selection rules is sensitive also to modes between the instanton and the $I_1$-locus of the discriminant, which is responsible for the $U(1)_X$.

Although we have considered specifically an $SU(5)\times U(1)$ model in the above, the general argumentation is independent of the specific F-theory model. We expect that the surfaces which contribute non-trivial selection rules via \eqref{sel_rule_gen_surface} are surfaces which correspond to $\mathbb P^1s$ fibered over curves of singularity enhancement in the base.

It is interesting to compare to situations without an extra $U(1)$ group, in this case to generic $SU(5)$ Tate models.
The surface $-D_M \wedge \tw_X$ ceases to exist as an algebraic surface for generic complex structure moduli. What takes its place \cite{Braun:2011zm} (see also \cite{Krause:2011xj,Krause:2012yh} in the $SU(5)$ context) is a 4-cycle that can be written as a complete intersection only on the ambient fivefold $X_5$, but not on $\hat Y_4$. Its dual 4-form is an element of $H^{2,2}_{\rm hor}(\hat Y_4)$.
This surface now leads to a selection rule due to the Freed-Witten anomaly similar to the one for the matter surfaces\footnote{Note that in this recombined case also $G_4^X$ is no longer in $H^{(2,2)}_{\rm ver} (\hat Y_4)$.}. The resulting selection rules are expected to be equivalent in form to the ones in (\ref{sel_rule_G^X_over_C^X},\ref{sel_rule_G^lambda_over_C^X}), even though no interpretation in terms of a $U(1)$ charge is possible if the $U(1)$ is higgsed. The relevant integrals can be worked out by adopting and extending the techniques in \cite{Braun:2011zm}, but we do not present this computation here.

%%%%%%%%%%%%%%%%%%%%%%%%%%%%%%%%%%%%%%%%%%%%%%%%%%%%%%%%%%%%%%%%%%%%%%%%%%%%%%%%%%%%%%%%%%%%%%%%%%%%%%%%%%%%%%%%%%%%%%%%%
\subsection{Comparison with selection rules of Type IIB and $U(1)$ charges}
\label{sec:selection_rules_IIB}
%%%%%%%%%%%%%%%%%%%%%%%%%%%%%%%%%%%%%%%%%%%%%%%%%%%%%%%%%%%%%%%%%%%%%%%%%%%%%%%%%%%%%%%%%%%%%%%%%%%%%%%%%%%%%%%%%%%%%

Having discussed selection rules for the absence of chiral charged zero modes from an F-theory perspective, let us now see how this is related to the more familiar Type IIB picture. 

In Type II orientifolds the intersection of a D-brane instanton and a spacetime-filling D-brane hosts chiral charged zero modes in the sector of boundary changing open strings \cite{Blumenhagen:2006xt,Ibanez:2006da,Florea:2006si}, which is straightforward to quantise. In the case at hand, the net number of charged modes between an instanton along the divisor $D_E$ on $X_3$ a stack of $N_A$ D7-braned along the divisor $D_A$ is given by the chiral index
\bea
\label{chiral_A_E}
\chi_{A,E} =   - \int_{D_A \cap D_E} \tilde{{\cal F}}_A -  {\tilde{\cal F}}_E.
\eea
This counts the chirality of states in representation $({{\bf \bar N}_A}, 1_E)$.
As in section \ref{fluxedIIBInst} we use the notation $\tilde{\cal F} = 2\pi\alpha' F - B$.
Note that apart from the brane flux ${\cal F}_A$ on $D_A$ also the instanton flux ${\cal F}_E$ contributes to this chiral index \cite{Grimm:2011dj}. Furthermore the contribution of the continuous $B_-$ moduli drops out of the chiral index as they appear in both $\tilde{{\cal F}}_A$ and $\tilde{{\cal F}}_E$. The situation is different for the discrete $B_+$ moduli in the case of an $O(1)$-instanton as the full orientifold-even part $\tilde{{\cal F}}^+_E$ is projected out. Therefore the chiral index for an $O(1)$-instanton depends on ${{\cal F}}_A^{\alpha}$ rather than ${\cal F}_A^{\alpha}$, matching the $U(1)$ charge of the instanton as discussed below.

Equivalently, the presence of a net number of chiral zero modes can be detected from the net charge of the instanton action under each $U(1)_A$.
For a stack of $N_A$ spacetime-filling D7-branes on a divisor $D_A$ and the image stack along $D_A'$ this works as follows. The Poincar\'e dual 2-forms are expanded as $[D_A] = \frac12 C_A^\alpha \omega_\alpha + \frac12 C^a_A \omega_a$ and $[D_A'] = \frac12 C_A^\alpha \omega_\alpha - \frac12 C^a_A \omega_a$.  
The St\"uckelberg mechanism then leads to a gauging of the shift symmetries of the scalars $c^a$ and $c_\alpha$, such that under a gauge transformation  $A^A \rightarrow A^A + \frac{1}{\ell_s^2} d \Lambda^A$\footnote{The explicit factor of $\frac{1}{\ell_s^2}$ has been included to account for the difference in normalisation between the gauge field $A^A$ in the type IIB limit and the gauge field $A^A$ appearing in the expansion of $C_3$ in \eqref{C3A}, see \cite{Grimm:2011tb}. This implies that the gauge parameter $\Lambda$ appearing here can be identified with the parameter appearing in \eqref{gauge_trf_A}.} of the diagonal $U(1)_A$ the chiral fields \eqref{GaTalpha} shift as
\bea
G^a &\rightarrow& G^a + \frac{1}{2\pi}N_A C^a_A \Lambda^A, \nonumber \\
T_\alpha &\rightarrow& T_\alpha - \frac{i}{2\pi} N_A\Bigl( \cK_{\alpha \beta \gamma}  \, \tilde{\cal F}_A^{\beta} \, C^{\gamma}_A  +   \cK_{\alpha b c }  \, {\cal F}_A^b   \, C^{c}_A \Bigr) \Lambda^A.
\eea
The partition function (\ref{ZE3_rewrite}) of the $O(1)$ E3-instanton with a fixed flux configuration then also shifts as 
\be
\label{charge_instanton_IIB}
 e^{-S_E} \rightarrow e^{  - i q_A \Lambda^A} e^{-S_E}, \qquad
q_A =   - \frac{1}{2}N_A \Big(  \, \kappa_{\alpha b c } \, C^{\alpha}_E  \, C^b_A \, ({\cal F}^c_A - {\cal F}^c_E) + \kappa_{\alpha \beta \gamma} C^\alpha_E \, C^\beta_A \, {\tilde{\cal F}}^\gamma_A  \Big).
\ee
Note that only the pullback instanton flux contributes to the instanton charge, not the variable flux ${\cal F}_E^\tv$. It is simple to check that the charge $q_A$ precisely returns the chiral index \eqref{chiral_A_E}. \\

As a consequence of the non-trivial $U(1)_A$ charge $q_A$ associated with a given instanton flux, each configuration with charge $q_A$
contributes now to a superpotential term of the form $W \simeq {\cal O} \, {\rm exp}(-S_E)$, where  ${\cal O} $ is an operator involving open string fields with $U(1)_A$  charge $- q_A$.
The sum over the instanton fluxes in the instanton partition function then splits into various contributions. Each such contribution involves the sum over the full allowed lattice of variable instanton fluxes ${\cal F}_E^\tv$ together with the sublattice of allowed pullback fluxes  whose induced charge matches that of the respective operator ${\cal O}$.

In order to compare the orientifold description with the results of the previous subsection we first recall from  \cite{Krause:2012yh}  the D7-brane and flux configuration of the Type IIB model corresponding to the F-theory setup with gauge group $SU(5)\times U(1)_X$.  
The IIB setup consists of a stack of 5 D7-branes on a divisor $D_A$ on $X_3$ together with its orientifold image $D_A'$ and one additional D7 on $D_B = 4 D_{O7} - 2 {D_A} - 3 {D'_A}$ along with the associated image stack. 
For the reader's convenience we reiterate that the F-theory base $B_3$ is the orientifold quotient of $X_3$, i.e. there is a projection $p: X_3 \rightarrow B_3$. The $SU(5)$ divisor $\cal W^{\rm b}$ and the instanton divisor $D^{\rm b}_M$   in $B_3$  are given as
\bea
D_A^+ = D_A+ D_A' = p^*{\cal W}^{\rm b} ,    \qquad     D_E = p^*D_M^{\rm b}.
\eea
The class $D_{O7}$ of the orientifold plane maps to the anti-canonical class of $B_3$, $\bar{\cal K} = p^*D_{O7}$. Finally mind the factor of $2$ in
\bea
\int_{X_3} p^*({D}^{\rm b}_i) \wedge  p^*({D}^{\rm b}_j) \wedge  p^*({D}^{\rm b}_k) = 2 \int_{B_3} {D}^{\rm b}_i \wedge  {D}^{\rm b}_j \wedge  {D}^{\rm b}_k.
\eea

The two Abelian gauge symmetries $U(1)_A$ and $U(1)_B$ on $D_A$ and $D_B$ are individually massive even in absence of gauge flux, but there is a massless linear combination
\be
\label{massless_u1}
U(1)_X = \frac12 (U(1)_A - 5\ U(1)_B),
\ee
which only becomes massive if gauge flux is turned on.
The Type IIB flux configuration corresponding to the $G_4$-fluxes defined in \eqref{G_4_X} and \eqref{G_4_lambda} is given by \cite{Krause:2012yh}
\bea
\label{fluxIIB_G_4^X}
G_4^X : \qquad & \tilde{\cF}^+_A = \frac{1}{10} \cF, \quad & \tilde{\cF}^+_B = - \frac12 \cF,\quad \cF^-_{A,B}=0, \\
G_4^\lambda : \qquad & \tilde{\cF}^+_A = \frac{2 }{5} \lambda D_{O7}, \quad & \tilde{\cF}^+_B = 0,\quad\quad\quad \cF^-_{A,B}=0  \label{fluxIIB_G_4^lambda}
\eea
with
\bea
{\cal F} = p^*{\cal F}^b.
\eea

In particular we may compute the charge of the E3-instanton under the massless $U(1)_X$ of \eqref{massless_u1} as
\be
\label{chirality_massless_u1}
q^X = \frac{1}{2} (5 \chi_{A,E} - 5 \chi_{B,E}).
\ee
The factor of 5 in front of $\chi_{A,E}$ is due to the fact that the zero modes at this intersection locus transform in the $\bar{\textbf{5}}$ of the $U(5)$ along $D_A$.

Let us consider first the flux configuration \eqref{fluxIIB_G_4^X} corresponding to $G_4^X$.
From the chiral indices
\bea
&&  \chi_{A,E}(G_4^X) = -  \frac{1}{20} \int_{X_3}  D_E \wedge D_A^+  \wedge {\cal F} =  - \frac{1}{10} \int_{B_3} D_M^{\rm b} \wedge  {\cal W}^{\rm b} \wedge {\cal F}^{\rm b}, \\
&&\chi_{B,E}(G_4^\lambda) =  \frac{1}{4} \int_{X_3} D_E \wedge D_B^+  \wedge {\cal F} =  \frac{1}{2} \int_{B_3} D_M^{\rm b} \wedge  (8 \bar{\cal K}- 5 {\cal W}^{\rm b}) \wedge {\cal F}^{\rm b}
\eea
one deduces the overall $U(1)_X$ charge 
\bea
\frac{1}{2}(5 \chi_{A,E} - 5 \chi_{B,E}) =  - 10 \int_{B_3} D_M^{\rm b} \wedge (  \bar{\cal K}  - \frac{3}{5} {\cal W}^{\rm b}  )   \wedge {\cal F}^{\rm b}.
\eea

This precisely reproduces the F-theory result \eqref{sel_rule_G^X_over_C^X}.

In both pictures the first necessary selection rule for generation of a superpotential without insertion of extra net charged matter is therefore that
\eqref{sel_rule_G^X_over_C^X}
\be
 \int_{B_3} D_M^{\rm b} \wedge ( \bar{\cal K}  - \frac{3}{5} {\cal W}^{\rm b} )   \wedge {\cal F}^{\rm b} \stackrel{!}{=} 0.
\label{sel_rule_G^X_u1x}
\ee
From the type IIB perspective it is clear that the absence of chiral charged zero modes actually requires the vanishing of the chiral indices $\chi_{A,E}$ and $\chi_{B,E}$ individually, not just in the combination \eqref{chirality_massless_u1}. The complementary condition can be obtained in IIB by demanding the vanishing of the net chiral index with respect to any other $U(1)$ linearly independent from \eqref{massless_u1}. However all other combinations of $U(1)_A$ and $U(1)_B$ are massive, and therefore not directly visible in the F-theory model \cite{Grimm:2011tb}. Therefore the corresponding selection rule in F-theory cannot be derived in the manner presented above for $U(1)_X$. Instead, it must be contained in the additional selection rules discussed in the previous subsection which were obtained by integrating $G_4$ over one of the matter surfaces. Indeed, for the flux $G_4^X$ the selection rule \eqref{sel_matter_surface} just corresponds to the vanishing of $\chi_{A,E}$. Together with \eqref{sel_rule_G^X_u1x} this suffices to guarantee the vanishing of each chiral index.

Put differently, the massive $U(1)$s, even though inivisible in the spectrum of the F-theory compactification, do remain as selection rules in F/M-theory.
They enter through the Freed-Witten anomaly, which enforces that $\iota^*G_4 =0$ on the M5-instanton. While the spacetime interpretation is different, the effect is the same in F-theory compared to Type IIB orientifolds. 

That the massive $U(1)$s would continue to play a role as selection rules was conjectured from a different perspective in \cite{Grimm:2011tb}. Formally, the massive $U(1)$s can also be made visible by including  a set of non-harmonic 2-forms and 3-forms in the dimensional reduction of the M-theory supergravity.
In this approach the gauging induced by the massive $U(1)$ is encoded in the intersection form of these non-harmonic forms, and a definition of the instanton charge along the lines of (\ref{U1chargea}) is possible by replacing $\tw_X$ with the non-harmonic 2-form associated with the massive $U(1)$. The results of the present analysis are in agreement with \cite{Grimm:2011tb} even though no use of the non-harmonic forms has been made.

For the flux configuration $G_4^\lambda$ the charged zero mode indices follow as
\bea
\chi_{A,E}(G_4^\lambda)= - \frac{ \lambda}{5} \int_{X_3} D_{O7} \wedge D_E \wedge D^+_A = - \frac{2 \lambda}{5}\int_{B_3} \bar{\cal K} \wedge D_M^{\rm b} \wedge {\cal W}, \quad \quad
\chi_{B,E}(G_4^\lambda)=0.
\eea

The resulting $U(1)_X$ charge
\bea
q^X= - \lambda \int_{B_3} \bar{\cal K} \wedge D_M^{\rm b} \wedge {\cal W}
\eea
is in agreement with the F-theory result (\ref{sel_rule_G^lambda_over_C^X}).
Since $\chi_{B,E}(G_4^\lambda)=0$, there is no further selection rule in Type IIB.
At first sight this may seem puzzling, as in the F-theory setup we obtained the two selection rules \eqref{sel_rule_G^lambda_over_C^X} and \eqref{sel_matter_surface}. To understand this seeming discrepancy, we must recall that a smooth Type IIB limit actually exists only if the conifold point \cite{Donagi:2009ra, Krause:2012yh}
\be
\label{conifold_point}
D_A \cap \{a_1 = 0 \} \cap \{ a_{2,1} = 0\}
\ee
is absent, with the polynomials $a_1$ and $a_{2,1}$ defined in (\ref{Tatepolys}).
Moreover we have assumed from the beginning that the $SU(5)$ divisor ${\cal W}^{\rm b}$ is smooth. 
Type IIB orientifolds corresponding to this class of models satisfy the relation \cite{Krause:2012yh}
\bea
(D_A^+)^2 - (D_A^-)^2 = 2 D^+_A D_{O7}. 
\eea
Expression \eqref{sel_matter_surface} therefore equals
\bea
\label{integralmatch}
\frac{\lambda}{5} \int_{B_3} D_M^{\rm b} \wedge {\cal W}^{\rm b} \wedge (6 \bar{\cal K} - 5 {\cal W}^{\rm b}) = - \frac{2 \lambda}{5} \int_{X_3} D_E \wedge D_A^+ \wedge D_{O7} - \frac{\lambda}{2} D_E \wedge (D_A^-)^2.
\eea

Next note that for a general divisor the intersection $D_A \cap D'_A$ splits into a component along the O7-plane and a component away from it. Cohomologically the latter is  proportional to $D_A^+ \wedge D_A^+|_{\rm off \,  O7} + D_A^- \wedge D_A^- - 2 D_A^- \wedge D_A^+ $. In a Type IIB orientifold corresponding to an F-theory model with smooth $SU(5)$ divisor ${\cal W}^{\rm b}$, the component away from the O7-plane is absent because this is the locus that would translate into self-intersections of ${\cal W}^{\rm b}$. Now,  $\int_{X_3} D_E \wedge D_A^- \wedge D_A^+ =0$ because $D_E$ is orientifold even; furthermore we argued that the intersection $D_M^{\rm b} \cap {\cal W}^{\rm b} $ can be deformed inside $a_1=0$, i.e. the integral $\int_{X_3} D_E \wedge D_A^+ \wedge D_A^+|_{\rm off \, O7}=0$. In conclusion the last term in (\ref{integralmatch}) vanishes.

Hence as long as there exists a smooth Type IIB limit we indeed obtain only one selection rule for this type of flux configuration, as expected. 
Note however that \eqref{conifold_point} is not required to be absent in a general F-theory model, and the number of independent selection rules may be higher in this case. In fact, the locus \eqref{conifold_point} describes a point of enhancement to $E_6$ \cite{Donagi:2009ra}, and one might speculate that new charged zero mode states can appear at this exceptional singularity. This intuition is supported by our observation that additional selection rules not encountered in Type IIB can appear in the presence of a point of exceptional enhancement inside the instanton divisor, but more work is needed to understand this microscopically.

We would like to close this section with an even more speculative remark:
For fluxed D3-instantons in Type IIB orientifolds, the chiral index (\ref{chiral_A_E}) counting the charged zero modes depends not only of the 7-brane fluxes, but also on the instanton flux $\tilde {\cal F}_{E}$. In view of (\ref{charge_instanton_IIB}) this effect induces an instanton flux contribution to the net $U(1)_X$ charge of the instanton and can in particular be used to cancel the chirality induced by D7-brane gauge flux \cite{Grimm:2011dj}. 
On the other hand, no analogous phenomenon has been observed for the M5-instanton flux quanta $\cal H$ in the present treatment.
A possible resolution of this mismatch is related to the fact that the relevant D3-instanton fluxes arise from 2-forms in $H_-^{1,1}(D_E)$ with non-vanishing pullback onto the D7-brane stacks. As indicated in section \ref{sec:O(1)Uplift}, the uplift of such 2-forms to the cohomology of $D_M$ is unclear.
Formally it is possible introduce a set of non-harmonic 3-forms on $D_M$ to capture the missing ${\cal H}$ fluxes such as to recover the $U(1)$ charges, in a similar spirit to the analysis in \cite{Grimm:2011tb}. It would be desirable to understand, however, whether a different and more concrete description of this effect is possible.
 We hope to return to this interesting point in the future.

%%%%%%%%%%%%%%%%%%%%%%%%%%%%%%%%%%%%%%%%%%%%%%%%%%%%%%%%%%%%%%%%%%%%%%%%%%%%%%%%%%%%%%%%%%%%%%
\section{Conclusions and  open questions}
\label{sec_concl}
%%%%%%%%%%%%%%%%%%%%%%%%%%%%%%%%%%%%%%%%%%%%%%%%%%%%%%%%%%%%%%%%%%%%%%%%%%%%%%%%%%%%%%%%%%%%%%

In this paper we have studied the superpotential due to M5-instantons in F-theory compactifications. The discussions are framed in a general F-theoretic setting, but are augmented, where possible, by comparison with known Type IIB results.

In the first part of the paper we have discussed the classical partition function of M5-instantons, focussing on instantons wrapped on vertical divisors of the elliptically fibered fourfold of F-theory. These can be related to E3-instantons in the corresponding Type IIB model.
 By comparing the results with the analogous expressions for the E3-instanton we have been able to identify the correct partition function of the M5-instanton, which amounts to choosing a specific line bundle on the intermediate Jacobian of the M5-instanton divisor. Our result shows that there is a canonical choice independent of the precise geometry of this divisor. The calculations required to extract the correct line bundle directly in M-theory are highly non-trivial, and it is not obvious that such a simple canonical choice should exist.
 The M5 partition function identified using the duality with Type IIB is in agreement with the results of \cite{Dolan:1998qk, Gustavsson:2000kr, Gustavsson:2011ur, Dijkgraaf:2002ac}, which calculate the partition function on a flat torus resp. on $K3\times T^2$. However, our result is applicable to a much larger class of instanton geometries; in particular the existence of a smooth Type IIB limit only poses constraints on the form of the discriminant locus, but not of the M5 divisor. Thus we expect this result to hold also if no such limit can be taken smoothly.

In the second part of this work we have focused on the question of how the presence of $G_4$-flux restricting non-trivially to the instanton affects the M5-instanton superpotential contribution. Our discussion is applicable to general $G_4$-fluxes and valid even in absence of a smooth Type IIB limit.
It is well-known that in Type IIB gauge fluxes can generate a chiral spectrum of charged fermionic zero modes on the E3-instanton. The presence of such chiral zero modes implies that the instanton can at most contribute to superpotential couplings involving charged open string fields. We have discussed the corresponding effect from the perspective of M-theory, where it has a close relationship with the Freed-Witten anomaly cancellation condition \cite{Donagi:2010pd, Marsano:2011nn}. This condition implies that a consistent theory of the M5-instanton in the presence of $G_4$-flux requires the presence of a suitable configuration of M2-branes ending on the instanton. We have argued that in the resulting superpotential contributions, these M2-branes play the role of the charged open string fields in the dual Type IIB picture. Related recent work on these instanton zero modes can be found e.g. in \cite{Blumenhagen:2010ja, Donagi:2010pd, Marsano:2011nn, Cvetic:2011gp}.

In many applications one is interested in generating superpotential contributions that do not involve charged operators. A necessary condition for this is encoded in the selection rule $\iota^* G_4 = 0$. This selection rule guarantees absence of chiral charged fermionic zero modes. Of course there can exist further neutral zero modes which forbid the generation of a superpotential term. These neutral zero modes have been discussed e.g. in \cite{Saulina:2005ve, Kallosh:2005gs, Martucci:2005rb, Blumenhagen:2010ja} and are not considered further in this work. Focussing on a specific set of non-Cartan $G_4$-fluxes that were recently constructed in  \cite{Marsano:2011hv, Krause:2011xj,Krause:2012yh,Grimm:2011fx}, we have discussed a set of explicit 4-cycles on the M5-instanton that can be used to numerically test for the condition $\iota^* G_4 = 0$ by evaluating the integral of $G_4$ over these cycles. Following this discussion in a general F-theoretic setting, we then proceed to illustrate these ideas in the context of the explicit $SU(5)\times U(1)$ model as considered in \cite{Krause:2012yh}. In particular we have checked that if a smooth Type IIB limit exists, the selection rules obtained in this manner correspond exactly to the vanishing of the chiral indices counting zero modes at the intersections between the E3-instanton and D7-brane stacks.

It is well known that in the Type IIB setting these selection rules for the absence of chiral charged zero modes can be obtained by considering the charge of the instanton under Abelian gauge symmetries present in the model. As is to be expected, we have found that a corresponding observation can be made in F-theory. Namely, in models including a massless $U(1)$ gauge symmetry the integral of $G_4$ over some of the above 4-cycles computes the instanton charge
under this massless $U(1)$ symmetry. Furthermore, we have shown that the additional selection rules which arise from the remaining types of surfaces can be interpreted as instanton charges under massive $U(1)$ gauge symmetries, confirming the expectations of \cite{Grimm:2011tb}.

There are several obvious directions for future study. 
We have performed the explicit match of the partition functions for the case of $O(1)$ E3-instantons wrapped on a rigid, irreducible divisor invariant under the orientifold action. A number of questions arise when one relaxes these conditions. If one considers a non-rigid divisor one obtains in Type IIB a non-trivial F-term condition restricting the instanton flux which can contribute to the partition function. In the dual F-theory setting one can immediately identify a corresponding set of M5-instanton fluxes which are expected to be ruled out by a similar condition. However, the obvious uplift to F-theory of the Type IIB F-term expression does not seem sufficient to rule out all of these fluxes, and it would be interesting to study how the full required constraint arises in F-theory. Another interesting question would be how to describe the uplift of Type IIB $U(1)$ instantons, which admit fluxes of even orientifold parity contributing to the partition function. Such fluxes uplift to 2-forms on the M5-instanton with no obvious interpretation as M5 3-form flux $\cH$.

Further points requiring further investigation concern the charged zero modes of the M5-instanton and the resulting selection rules.
While we have discussed in general terms how the surfaces in the M5-instanton which can be used to detect non-trivial selection rules should arise, the construction of these surfaces in the framework of the model considered in section \ref{sec_su5xu1} relied on a specific technical assumption. While it is clear that the technical assumption in question is not a necessary condition for the existence of the surfaces, it would be interesting to see how the proof can be generalised.
In addition, it would be very nice to understand in more detail how the charged zero modes are to be understood in terms of M2-branes ending on the instanton and in particular how the appearance of these zero modes is related to the geometry of the singular elliptic fiber over an intersection with a D7-brane stack. A precise understanding of the necessary geometric requirements to give rise to such M2-brane states should also allow one to understand the role of vectorlike pairs of zero modes. Another approach would be to use more extensively the correspondence between M2s ending on the M5-instanton and non-trivial $\cH$-flux configurations accompanied by a suitable deformation of the M5 in the M2-brane directions. This would allow one to extract information on the zero modes by studying instead the zero modes of this $\cH$-field configuration.

Finally, it would be interesting to consider how non-zero $\cH$-flux affects the instanton selection rules. In Type IIB theory it is known that instanton fluxes which restrict non-trivially to the intersection of instanton and brane contribute to the chiral index or equivalently the $U(1)$ charge of the instanton \cite{Grimm:2011dj}. However, the uplift of precisely this type of fluxes to F-theory is problematic due to the degeneration of the elliptic fiber over the location of the D7-branes. In \cite{Grimm:2011tb} it was conjectured that 2-forms with negative orientifold parity restricting to the D7-branes uplift to non-harmonic 3-forms in F-theory. Following this idea suggests that IIB instanton flux that contributes to the chiral index should be described by non-harmonic $\cH$-flux on the M5-instanton. Intriguingly, one finds exact match with the Type IIB expression if one computes the $U(1)$ charge of the M5-instanton after including $\cH$-flux along the non-harmonic forms introduced in \cite{Grimm:2011tb}. However, it is so far unclear how the appearance of non-harmonic $\cH$-flux is to be understood from a purely M-theoretic perspective. We look forward to revisiting these and related question in the future.

%%%%%%%%%%%%%%%%%%%%%%%%%%%%%%%%%%%%%%%%%%%%%%%%%%%%%%%%%%%%%%%%%%%%%%%%%%%%%%%%%%%%%%%%%%%%%%
\subsection*{Acknowledgements}
%%%%%%%%%%%%%%%%%%%%%%%%%%%%%%%%%%%%%%%%%%%%%%%%%%%%%%%%%%%%%%%%%%%%%%%%%%%%%%%%%%%%%%%%%%%%%%
We thank Christoph Mayrhofer for very helpful discussions. We are grateful to Thomas Grimm and Eran Palti for collaboration on the related \cite{Grimm:2011dj,Grimm:2011tb}. MK acknowledges support by the IMPRS for Precision Tests of Fundamental Symmetries and the German National Academic Foundation. Special thanks to the Simons Center for Geometry and Physics for support and hospitality. This work was funded in part by the DFG under Transregio TR 33 "The Dark Universe".

%%%%%%%%%%%%%%%%%%%%%%%%%%%%%%%%%%%%%%%%%%%%%%%%%%%%%%%%%%%%%%%%%%%%%%%%%%%%%%%%%%%%%%%%%%%%%%
\appendix

\section{M-theory reduction in the democratic formulation}
\label{democratic_reduction}
The low-energy limit of M-theory is conventionally described by the action
\be
\label{S11normal}
S = \frac12 \int d^{11} x \sqrt{-g} R - \frac{1}{4} \int \left( G_4 \wedge \ast G_4 + \frac{1}{3} C_3 \wedge G_4\wedge G_4 \right)
\ee
in conventions where $\kappa_{11}^2 = 1$. The dimensional reduction to three dimensions of the massless harmonic sector of this action on a Calabi-Yau fourfold as well as its F-theory limit has been extensively studied  e.g. in \cite{Haack:1999zv, Haack:2001jz, Grimm:2010ks}. Furthermore in \cite{Grimm:2011tb} it was discussed how certain non-harmonic forms can be included in the dimensional reduction to describe massive $U(1)$ symmetries. However, in order to describe the interaction between M5-instantons and the $C_3$ field it is advantageous to stray from this conventional path and use instead a democratic formulation which includes also the magnetic dual $C_6$ of $C_3$. This mirrors the situation in Type II string theory, where one uses the democratic formulation upon including D-branes.

The duality-symmetric eleven-dimensional supergravity theory has been considered by the authors of \cite{Bandos:1997gd}, who find that one needs to introduce an additional auxiliary scalar field to obtain a manifestly covariant theory. In the following we will not follow this approach; instead we will describe the theory by means of a simpler pseudo-action which is supplemented by a set of duality constraints to be imposed at the level of the equations of motion. This is again analogous to the approach conventionally followed in the setting of Type II string theory.

In the theory defined by \eqref{S11normal} the dynamics of $C_3$ are determined by the equation of motion
\be
\label{eom_c3}
d \ast G_4 + \frac12 G_4\wedge G_4 = 0
\ee
as well as the Bianchi identity $dG_4 = 0$.
In close analogy to the democratic formulation of Type II string theory one can define a dual field strength
\be
G_7 = d C_6 + \frac12 C_3 \wedge G_4
\ee
and consider the action 
\be
\label{S11democ}
S_D = \frac12 \int d^{11} x \sqrt{-g} R - \frac{1}{6} \int \left( G_4 \wedge \ast G_4 + \frac12 G_7 \wedge\ast G_7 \right).
\ee
Supplementing this pseudo-action with the duality relation
\be
\label{duality_11D}
\ast G_7 = G_4 \quad \Leftrightarrow \quad G_7 = - \ast G_4
\ee
one easily checks that it reproduces precisely the original equations of motion \eqref{eom_c3} as well as the relevant Bianchi identities. The prefactor of the second term can be fixed by requiring that integrating out $C_6$ by introducing a Lagrangian multiplier term $-\int dC_6 \wedge dC_3$ leads back to the original action \eqref{S11normal}. While the form of the action \eqref{S11democ} is well known, to the best of our knowledge its dimensional reduction to three dimensions has not been performed in the literature. This will be the subject of this appendix.

The advantage of using the democratic formulation when working with M5-instantons is that it explicitly involves the field $C_6$ which couples directly to the instanton world-volume. This coupling involves a set of scalars $c_\alpha$ which as we will see below become part of the complexified K\"ahler moduli. Performing the dimensional reduction of \eqref{S11democ} in the democratic formulation allows us to determine the correct normalisation of the K\"ahler moduli. This in particular allows us to compute the dependence of the M5-instanton action on the K\"ahler moduli including all prefactors and to perform the precise match with the Type IIB expression.
 
It is worth noting that the dimensional reduction of the usual action \eqref{S11normal} yields a three-dimensional theory involving vectors $A^\alpha$ which can in three dimensions be dualised to scalars $\tilde{c}_\alpha$\cite{Haack:1999zv, Haack:2001jz}. However even if one does this and rewrites the bulk action in terms of these scalars $\tilde{c}_\alpha$, one does not know precisely how the so obtained $\tilde{c}_\alpha$ should appear in the instanton action. In fact as we will see there is a non-trivial shift between the fields appearing naturally in the M5 action and the fields that form part of the K\"ahler coordinates, which one can only determine precisely when performing the reduction in the democratic formulation.

To carry out the dimensional reduction of the action \eqref{S11democ} let us first focus only on the massless fields described by harmonic forms and introduce the bases $\{ \omega_{\Lambda} \}$ resp. $\{ \tilde{\omega}^{\Lambda} \}$ of $H^{1,1}(Y_4)$ resp. $H^{3,3}(Y_4)$, $\{ \eta^m \}$ of $H^4 (Y_4)$ and $\{\alpha_a, \beta^b \}$ resp. $\{ \tilde{\alpha}^a , \tilde{\beta}_b \}$ of $H^3 (Y_4)$ resp. $H^5(Y_4)$. These forms are chosen to obey
\bea
\int \omega_\Lambda \wedge \tilde{\omega}^\Sigma = \frac12 \delta_\Lambda^\Sigma, && \Lambda, \Sigma = 1, \ldots , h^{1,1}(Y_4), \\
\int \eta^m \wedge \eta^n = \delta^{mn}, && m, n = 1, \ldots, h^4(Y_4), \\
\genfrac{.}{\}}{0pt}{0}{\int \alpha_a \wedge \tilde{\alpha}^b = \int \beta^b \wedge \tilde{\beta}_a = \frac12 \delta_a^b \quad}{\int \alpha_a \wedge \tilde{\beta}_b = \int \beta^a \wedge \tilde{\alpha}^b = 0 \quad } && a,b = 1 ,\ldots, h^3(Y_4).
\eea
Recall that $H^{1,1}(Y_4)$ can be split into several distinct subspaces, namely the forms $\omega_\alpha$ which constitute the uplift of $H^{1,1}(B_3)$, the form $\omega_0$ which is Poincar\'e dual to the base $B_3$ as well as a set of resolution divisors $\omega_{I}$ describing massless $U(1)$ gauge symmetries\footnote{We do not need to distinguish at this point between forms describing $U(1)$s in the Cartan of a non-Abelian gauge group and additional Abelian gauge group factors. In this appendix $Y_4$ is always taken to denote the smooth, resolved fourfold.}. To simplify the notation let us further assume that $h^{1,2}(B_3)=0$,  so that we have the intersection numbers \cite{Grimm:2011tb}
\bea
\label{intersection_numbers_fourfold}
\int \omega_\Lambda\wedge\alpha_a\wedge\beta^b & = & \genfrac{\{}{.}{0pt}{0}{\frac12 \KK_{\alpha a c}\delta^{cb}, \qquad \Lambda = \alpha}{0, \qquad\qquad \Lambda = 0, I}, \\
\int \omega_\Lambda \wedge\alpha_a\wedge\alpha_b = \int \omega_\Lambda \wedge \beta^a \wedge\beta^b & = & 0, \\
\int \omega_\Lambda \wedge\omega_\Sigma \wedge \omega_\Pi \wedge \omega_\Theta & = & \frac12 \KK_{\Lambda\Sigma\Pi\Theta}.
\eea
Recall that in the case where the model admits a Type IIB limit on the double cover $X_3$ of $B_3$ the $\alpha_a$ and $\beta^b$ are the uplifts of $H^{1,1}_-(X_3)$. In this case the intersection numbers $\KK_{\alpha a b}$ and $\cK_{\alpha\beta\gamma} \equiv \cK_{0 \alpha\beta\gamma}$ can also be identified with the corresponding intersection numbers on $X_3$. However we reiterate that the existence of a smooth orientifold limit is of course not necessary for this reduction.

For further convenience let us rewrite the intersection numbers above in terms of the following equivalent set of equations in cohomology
\be
\label{cohom_identities}
\begin{array}{rclrcl}
 \omega_\Lambda \wedge\omega_\Sigma \wedge \omega_\Pi & = &  \KK_{\Lambda\Sigma\Pi\Theta} \tilde{\omega}^\Theta, \qquad & \alpha_a\wedge\beta^b & = & \KK_{\alpha a c} \delta^{cb} \tilde{\omega}^\alpha, \\
\alpha_a\wedge \omega_\alpha & = & - \KK_{\alpha a c } \delta^{cb}\tilde{\beta}_b, \qquad & \beta^b \wedge \omega_\alpha & = & \KK_{\alpha a c} \delta^{cb} \tilde{\alpha}^a, \\
\alpha_a \wedge\alpha_b & = 0 = & \beta^a\wedge\beta^b, \qquad & \alpha_a\wedge\omega_\Lambda & = 0 = & \beta^a\wedge\omega_\Lambda, \quad \Lambda = 0,\ I, \\
\omega_\Lambda\wedge\omega_\Sigma & \equiv & R_{m \Lambda\Sigma } \eta^m & \Rightarrow \omega_\Lambda \wedge \eta^m & = & 2 \delta^{mn} R_{n \Lambda \Sigma} \tilde{\omega}^{\Sigma}.
\end{array}
\ee

We may now expand $C_3$ and $C_6$ into the bases of the cohomology groups defined above as follows
\bea
\label{expansions_C3_C6}
C_3 &=& A^\Lambda \wedge \omega_\Lambda + c^a \alpha_a + b_b \beta^b, \\
C_6 &=& \tilde{c}_\Lambda \tilde{\omega}^\Lambda + \tilde{U}_a \wedge \tilde{\alpha}^a + \tilde{V}^b \wedge\tilde{\beta}_b + r_m \wedge \eta^m.
\eea
Here $c^a,\ b_b,\ \tilde{c}_\Lambda$ are scalars, while $A^\Lambda,\ U_a,\ V^b$ are vectors and the $r_m$ are spacetime 2-forms, which are non-dynamical in three dimensions. We also include non-trivial $G_4$-flux
\be
G_4 = d C_3 + \cF_m \eta^m,
\ee
which is dual to the 2-forms $r_m$. We are now in a position to insert these expansions into the action \eqref{S11democ} and integrate out the internal space $Y_4$. To simplify the resulting expression let us finally introduce the notation $\{ \rho_i \} \equiv \{\alpha_a, \beta^b \}$, $ \{ \tilde{\rho}^i \} = \{ \tilde{\alpha}^a, \tilde{\beta}_b \}$ and
\bea
\label{def_N}
N_{\Lambda\Sigma} := \int \omega_\Lambda \wedge\ast \omega_\Sigma  & \Leftrightarrow & \int \tilde{\omega}^\Lambda \wedge\ast \tilde{\omega}^\Sigma = \frac{1}{4} N^{\Lambda\Sigma}, \\
M_{ij} := \int \rho_i \wedge \ast \rho_j  & \Leftrightarrow & \int \tilde{\rho}^i \wedge \ast \tilde{\rho}^j = \frac{1}{4} M^{ij}, \\
P^{nm} := \int \eta^n \wedge\ast \eta^m,  && 
\eea
where $N^{\Lambda\Sigma}$ denotes the inverse matrix of $N_{\Lambda\Sigma}$ etc.. Ignoring the Einstein-Hilbert term in \eqref{S11democ} for the moment one finds the following action for the fields arising from $C_3$ and $C_6$
\bea
\label{S_redux_1}
 S_C &= -\frac{1}{6} \int_{\mathbb{R}^{1,2}} \Bigl[ & N_{\Lambda\Sigma} F^\Lambda\wedge\ast F^\Sigma + M_{ij} dc^i\wedge\ast dc^j + \frac{1}{8} N^{\Lambda\Sigma} \cD c_\Lambda \wedge \ast \cD c_\Sigma \\
&& + \frac{1}{8}M^{ij} \nabla U_i \wedge\ast \nabla U_j + \frac12 P^{mn} \cD r_m \wedge \ast \cD r_n + P^{mn} \cF_m \cF_n \ast 1 \Bigr]. \nonumber
\eea
The shifted fields and covariant derivatives appearing here are given by
\be
\begin{array}{rcll}
c_\Lambda &=& \genfrac{\{}{.}{0pt}{0} {\tilde{c}_{\Lambda} ,}{ \tilde{c}_\alpha + \frac12 \KK_{\alpha a c}\delta^{cb} c^a b_b,} & \genfrac{.}{.}{0pt}{0} {\Lambda = 0,\ I}{\Lambda = \alpha},   \\
\cD c_{\Lambda} & := & dc_\Lambda + \frac12 \Theta_{\Lambda\Sigma}A^\Sigma + \genfrac{\{}{.}{0pt}{0} {0  ,}{ - \KK_{\alpha a c}\delta^{cb} c^a d b_b ,} & \genfrac{.}{.}{0pt}{0} {\Lambda = 0,\ I}{\Lambda = \alpha},  \\\\
U_i & = & \genfrac{\{}{.}{0pt}{0} {U_a - \frac12 \KK_{\alpha a c}\delta^{cb}b_b A^\alpha ,}{ V^b + \frac12 \KK_{\alpha a c}\delta^{cb} c^a A^\alpha ,} & \genfrac{.}{.}{0pt}{0} {U_i = U_a }{U_i = V^b }, \\
\nabla U_i & := & d U_i +  \genfrac{\{}{.}{0pt}{0} {\KK_{\alpha a c}\delta^{cb}b_b F^\alpha ,}{ - \KK_{\alpha a c}\delta^{cb} c^a F^\alpha ,} & \genfrac{.}{.}{0pt}{0} {U_i = U_a }{U_i = V^b }, \\\\
\cD r_m & := & d r_m + \frac12 R_{m \Lambda\Sigma} A^\Lambda \wedge F^\Sigma. &
\end{array}
\ee
The flux-dependent quantity $\Theta_{\Lambda\Sigma}$ controlling the gauging of the scalars $c_\Lambda$ is given by
\be
\label{def_Theta}
\Theta_{\Lambda\Sigma} = 2 \cF_m \delta^{mn} R_{n \Lambda\Sigma} = 2 \int \omega_\Lambda \wedge \omega_\Sigma \wedge G_4.
\ee

The next step is to eliminate the redundant degrees of freedom using the duality relation \eqref{duality_11D}, which after dimensional reduction can be written in terms of the three-dimensional fields as\footnote{Here we used the fact that due to the block-diagonal form of the metric one has $\ast_{11} (\omega \wedge \eta) = (-1)^{q(3-p)} (\ast_{3} \omega)\wedge (\ast_{Y_4} \eta)$ for a p-form $\omega$ on $\mathbb{R}^{1,2}$ and a q-form $\eta$ on $Y_4$; this was also used in deriving \eqref{S_redux_1}.}
\bea
\label{duality_1}
2 M_{ij} \ast dc^j = - \nabla U_i & \Leftrightarrow & \ast \nabla U_i = 2 M_{ij}d c^j, \\
\label{duality_2}
2N_{\Lambda\Sigma} \ast F^\Sigma = - \cD c_\Lambda & \Leftrightarrow & \ast \cD c_\Lambda = 2N_{\Lambda\Sigma} \ast F^\Sigma, \\
\label{duality_3}
- \delta_{mn} P^{np} \cF_p \ast 1 = \cD r_m & \Leftrightarrow & \cF_m = \delta_{mn} P^{np} \ast \cD r_p.
\eea
However, when computing the equations of motion of the fields $A^\Lambda$ arising from \eqref{S_redux_1} one finds that they are inconsistent with the duality relation given above. This phenomenon is a result of using the pseudo-action \eqref{S11democ} instead of the fully-fledged covariant action of \cite{Bandos:1997gd} (see also \cite{Pasti:1996vs}). It is also familiar from reductions of democratic Type II supergravity, compare e.g.  \cite{Dall'Agata:2001zh, Jockers:2004yj, Kerstan:2011dy, Grimm:2011dx} for discussions in the Type II setting. In order to restore consistency one must shift the definition of the field strengths $\cD c_\Lambda$ to
\be
\cD c_{\Lambda}  :=  dc_\Lambda + \Theta_{\Lambda\Sigma}A^\Sigma + \genfrac{\{}{.}{0pt}{0} {0  ,}{ - \KK_{\alpha a c}\delta^{cb} c^a d b_b ,}  \genfrac{.}{.}{0pt}{0} {\Lambda = 0,\ I}{\Lambda = \alpha}.
\ee
After adding a Lagrangian multiplier term $\frac{1}{6} \int \delta^{mn} \cF_n d r_m $ to the action one can treat $d r_m$ as an independent field and finds that its e.o.m. precisely give the duality relation \eqref{duality_3}. Integrating out $d r_m$ using these e.o.m. leads to
\bea
\label{S_redux_2}
 S_C &= -\frac{1}{6} \int_{\mathbb{R}^{1,2}} \Bigl[ & N_{\Lambda\Sigma} F^\Lambda\wedge\ast F^\Sigma + M_{ij} dc^i\wedge\ast dc^j + \frac{1}{8} N^{\Lambda\Sigma} \cD c_\Lambda \wedge \ast \cD c_\Sigma \\
&& + \frac{1}{8}M^{ij} \nabla U_i \wedge\ast \nabla U_j + \frac14 \Theta_{\Lambda\Sigma}A^\Lambda \wedge F^\Sigma + \frac{3}{2}P^{mn} \cF_m \cF_n \ast 1 \Bigr]. \nonumber
\eea
Next one could eliminate either the vectors $A^\Lambda$ or the scalars $c_\Lambda$ in favor of their dual fields leading to two different but equivalent formulations of the low-energy theory. As we want to study M5-instantons on vertical divisors of the form $[D_M] = C^\alpha_E \omega_\alpha$ we would like to keep the scalars $c_\alpha$ which enter the instanton action via the coupling $S_{M5} \supset 2\pi i \int_{D_M} C_6$. On the other hand we would like to keep $A^0$ resp. $A^{i}$ in the theory to facilitate the match with the Type IIB theory, where they correspond to the off-diagonal components of the 4d metric resp. the massless $U(1)$ gauge symmetries \cite{Grimm:2010ks}. This turns out to be possible if we assume that 
\be
\Theta_{\alpha\beta} = 0,
\ee
such that $\cD c_\alpha$ does not involve the dual degrees of freedom $A^\alpha$. In fact, $\Theta_{\alpha\beta} \propto \int \omega_\alpha\wedge \omega_\beta \wedge G_4 = 0$ is anyway required in order to obtain a consistent F-theory limit \cite{Grimm:2010ks, Grimm:2011tb}.
Letting the indices $r, s, ...$ run over the values $0, I$ here and in the following we again add a Lagrangian multiplier $\frac{1}{6} \int \frac12 d c_r \wedge F^r$ to the action. From the definition \eqref{def_N} and the different origins of the forms $\omega_\alpha$, $\omega_r$ one infers that the submatrices $N_{(1)} := (N_{\alpha\beta})$ and $N_{(2)} := (N_{rs})$ are separately invertible and positive definite. This implies that the same statement holds for their respective Schur complements $S_{(i)},\ i=1,2$, which are the inverses of the corresponding subblocks of the \emph{inverse} matrix $N^{\Lambda\Sigma}$, i.e. $S_{(1)rs} N^{sp} = \delta_r^p$, $S_{(2)\alpha\beta} N^{\beta\gamma} = \delta_\alpha^\gamma$. Therefore the e.o.m. 
\be
F^r = \frac12 N^{r \Lambda} \ast \cD c_\Lambda
\ee
obtained by treating $d c_r$ as an independent field suffices to completely determine $\cD c_s$ in terms of $F^r$ and $\cD c_\alpha$ and can therefore be used to integrate it out.

A similar argument shows that one can consistently eliminate $F^\alpha$ and $U_i$ after adding corresponding Lagrangian multiplier terms $-\frac{1}{6} \int F^\alpha \wedge d \tilde{c}_\alpha - \frac12 d U_i \wedge dc^i$. The final result of this dualisation can be concisely written in terms of the inverse $(N_{(1)}^{\alpha\beta})$ of $N_{(1)}$ and the  Schur complement $S_{(1)rs} = N_{rs} - N_{r\alpha}N_{(1)}^{\alpha\beta} N_{\beta s}$ and reads\footnote{Here we used the matrix inversion lemma which implies $N^{\alpha\beta}= N_{(1)}^{\alpha\beta} + N^{\alpha r} S_{(1)rs}N^{s\beta}$.}
\bea
\label{S_redux_3}
 S_C &= -\frac{1}{4} \int_{\mathbb{R}^{1,2}} \Bigl[ & S_{(1)rs} F^r\wedge\ast F^s - N_{(1)}^{\alpha\beta} N_{\beta r} \cD c_\alpha \wedge F^r   + \frac{1}{4} N_{(1)}^{\alpha\beta} \cD c_\alpha \wedge \ast \cD c_\beta \\
&& + M_{ij} dc^i\wedge\ast dc^j + \frac18 \Theta_{rs}A^r \wedge F^s + P^{mn} \cF_m \cF_n \ast 1 \Bigr]. \nonumber
\eea
The vectors $A^r$ and scalars $c_{\alpha}$ combine with scalars $v^\Lambda$ arising in the expansion of the K\"ahler form $J = v^\Lambda \omega_\Lambda$ into three-dimensional vector and scalar multiplets. These scalars enter the effective action through the reduction of the curvature scalar, which was performed in \cite{Haack:1999zv}. After combining the result with \eqref{S_redux_3} and performing a Weyl rescaling $g_{\mu\nu}\rightarrow \cV^{-2} g_{\mu\nu}$ with $\cV$ the volume of $Y_4$ one obtains
\bea
\label{S_redux_full}
 S_D &=& \frac12 \int_{\mathbb{R}^{1,2}}d^3x \sqrt{-g} R  -\frac{1}{4} \int_{\mathbb{R}^{1,2}} \Bigl[ \cV S_{(1)rs} (d\xi^r\wedge\ast d\xi^s + F^r\wedge\ast F^s)  \\ && - N_{(1)}^{\alpha\beta} N_{\beta r} \cD c_\alpha \wedge F^r 
 + \frac{1}{4\cV} N_{(1)}^{\alpha\beta} \cD c_\alpha \wedge \ast \cD c_\beta + \frac{1}{\cV} N_{(1)}^{\alpha\beta}d \cV_\alpha\wedge \ast d\cV_\beta \nonumber \\ &&
  + 4 G_{M\bar{N}}d z^M \wedge\ast d\bar{z}^{\bar{N}}  + \frac{1}{\cV} M_{ij} dc^i\wedge\ast dc^j 
 + \frac18 \Theta_{rs}A^r \wedge F^s + \frac{1}{\cV^3} P^{mn} \cF_m \cF_n \ast 1 \Bigr]. \nonumber
\eea
Here we have defined $\xi^r = \frac{v^r}{\cV}$ and
\be
\cV_{\Lambda} = \partial_{v^\Lambda} \cV = \frac{1}{3!}\int \omega_\Lambda\wedge J^3.
\ee
Although they will not be needed in the following the $h^{3,1}(Y_4)$ complex structure moduli $z^M$ are included for completeness. Given a basis $\{ \Phi_M \}$ of $H^{3,1}(Y_4)$ they are defined by $\delta g_{\bar{i}\bar{j}} = - \frac{1}{3|\Omega|^2}\bar{\Omega}_{\bar{i}}^{klm} (\Phi_M)_{klm \bar{j}}$, while the metric is given by
\be
G_{M\bar{N}} = \frac{\int \Phi_M \wedge \bar{\Phi}_{\bar{N}}}{\int \Omega\wedge\bar{\Omega}}.
\ee

The action \eqref{S_redux_full} can be put into the general form of a gauged three-dimensional $\cN=2$ supergravity as given in \cite{Grimm:2011tb}. To do this one combines the scalars $c^a$ and $b_b$ into complex scalar fields $N^a = c^a -i f^{ab}b_b$ with $f^{ab}$ a suitable function of the complex structure moduli \cite{Haack:1999zv, Grimm:2011tb} and defines the complexified K\"ahler moduli
\be
t_{\alpha} = 2 \cV_{\alpha} + i c_\alpha = \frac{1}{3!}\cK_{\alpha\Lambda\Sigma\Pi}v^\Lambda v^\Sigma v^\Pi + i c_\alpha.
\ee
If one performs the F-theory limit of vanishing fiber volume $\epsilon$ as in \cite{Grimm:2011tb} one finds that the fields $t_{\alpha}$ scale at zeroth order in $\epsilon$ as
\be
t_{\alpha} \longrightarrow \frac12 \cK_{\alpha\beta\gamma} v_B^\beta v_B^\gamma + i c_\alpha,
\ee
where $v_B^\alpha$ are the K\"ahler moduli of the base $B_3$ and $\cK_{\alpha\beta\gamma} \equiv \cK_{0 \alpha\beta\gamma}$ are the corresponding intersection numbers. If a Type IIB limit exists, comparing with the usual expressions shows in particular that the fields $c_\alpha$ can be identified with the corresponding axions appearing in the Type IIB K\"ahler moduli $T_\alpha$.

Finally let us consider the modifications that arise upon inclusion of the non-closed forms $\tw_{0A}$ introduced in \cite{Grimm:2011tb} to describe geometrically massive U(1) gauge fields. One change is that the indices $r,\ s,...$ introduced above now run also over these additional values. To stick to the notation of \cite{Grimm:2011tb} we introduce one such form $\tw_{0A}$ for each stack $A$ of 7-branes and let the index $i$ in $ r,s,... = 0, i, 0A$ now run only over all Cartan $U(1)$s. Note that although the $\tw_{0A}$ are individually non-closed and describe the massive diagonal $U(1)$ on each stack of D7-branes, certain linear combinations may remain massless, as is the case in the $SU(5)\times U(1)$ model discussed in section \ref{sec:selection_rules_Ftheory}. 

The non-closed forms induce a geometric gauging of the shift symmetries of the scalars $c^a$ resulting in the appearance of covariant derivatives
\be
\label{geometric_gauging}
d c^i \longrightarrow \nabla c^i = \genfrac{\{}{.}{0pt}{}{d c^a - N_{A}C^a_A A^{0A}, \qquad c^i = c^a.}{d b_b, \qquad c^i = b_b.}
\ee
Taking into account \cite{Grimm:2011tb}
\be
\omega_\Lambda \wedge d\beta^b = \genfrac{\{}{.}{0pt}{}{-\delta^{bc}\cK_{\alpha a c}N_A C^a_A \tilde{\omega}^\alpha , \qquad \text{if } \Lambda = 0A}{0, \qquad \text{otherwise}}
\ee
one also finds an additional piece in the covariant derivarives
\be
\cD c_{\alpha}  :=  dc_\alpha + \Theta_{\alpha \Sigma}A^\Sigma  - \KK_{\alpha a c}\delta^{cb} c^a d b_b -\frac12 \cK_{\alpha a c}\delta^{cb} b_b N_A C^a_A A^{0A}.
\ee
This implies that the shift symmetries of the scalars $c^a$, $\tilde{c}_\alpha$ are gauged, such that under a gauge transformation $A^{0A}\rightarrow A^{0A} + d\Lambda^A$ of the diagonal $U(1)$ the fields shift as
\be
\label{shifts_under_U1_A}
\delta_{\Lambda^A} c^a = N_A C_A^a \Lambda^A , \qquad \quad \delta_{\Lambda^A} \tilde{c}_\alpha = \frac12 \cK_{\alpha a c}\delta^{cb} b_b N_A C_A^a \Lambda^A -  \Theta_{\alpha  0A} \Lambda^A .
\ee
This implies that the shifted field $c_\alpha$ is gauged only in the presence of fluxes as $\delta_{\Lambda^A} c_\alpha = -  \Theta_{\alpha  0A} \Lambda^A$, as was to be expected from the Type IIB perspective.

Note that even if one picks fixed harmonic representatives for the basis forms above the equations \eqref{cohom_identities} still only hold in cohomology, as the wedge product of two harmonic forms is not necessarily harmonic. Upon inclusion of the non-closed $\tw_{0A}$ the additional exact pieces implicit in \eqref{cohom_identities} can lead to additional terms in the low-energy action. It is a non-trivial assumption that the intersection properties of the $\tw_{0A}$ are such that no such additional terms arise.

%%%%%%%%%%%%%%%%%%%%%%%%%%%%%%%%%%%%%%%%%%%%%%%
%%%%%%%%%%%%%%%%%%%%%%%%%%%%%%%%%%%%%%%%%%%%%%%
%%%%%%%%%%%%%%%%%%%%%%%%%%%%%%%%%%%%%%%%%%%%%%%
%%%%%%%%%%%%%%%%%%%%%%%%%%%%%%%%%%%%%%%%%%%%%%%
%%%%%%%%%%%%%%%%%%%%%%%%%%%%%%%%%%%%%%%%%%%%%%%
%%%%%%%%%%%%%%%%%%%%%%%%%%%%%%%%%%%%%%%%%%%%%%%
%%%%%%%%%%%%%%%%%%%%%%%%%%%%%%%%%%%%%%%%%%%%%%%
%%%%%%%%%%%%%%%%%%%%%%%%%%%%%%%%%%%%%%%%%%%%%%%

\clearpage
\bibliography{rev}  
\bibliographystyle{utphys}

%%%%%%%%%%%%%%%%%%%%%%%%%%%%%%%%%%%%%%%%%%%%%%%
%%%%%%%%%%%%%%%%%%%%%%%%%%%%%%%%%%%%%%%%%%%%%%%
%%%%%%%%%%%%%%%%%%%%%%%%%%%%%%%%%%%%%%%%%%%%%%%
%%%%%%%%%%%%%%%%%%%%%%%%%%%%%%%%%%%%%%%%%%%%%%%
%%%%%%%%%%%%%%%%%%%%%%%%%%%%%%%%%%%%%%%%%%%%%%%
%%%%%%%%%%%%%%%%%%%%%%%%%%%%%%%%%%%%%%%%%%%%%%%
%%%%%%%%%%%%%%%%%%%%%%%%%%%%%%%%%%%%%%%%%%%%%%%
%%%%%%%%%%%%%%%%%%%%%%%%%%%%%%%%%%%%%%%%%%%%%%%

\end{document}